\documentclass[]{aa}  

\usepackage{graphicx}
\usepackage{booktabs}
\usepackage{txfonts}
\usepackage{multicol}
\begin{document}

   \title{Radio continuum tails in ram pressure-stripped spiral galaxies: experimenting with a semi-empirical model in Abell 2255}
\titlerunning{Radio continuum tails in ram pressure-stripped spiral galaxies}

   \author{A. Ignesti
          \inst{1}
          \and
          B. Vulcani\inst{1}
          \and
          A. Botteon\inst{2}
          \and
          B. Poggianti\inst{1}
          \and
          E. Giunchi\inst{1,3}
          \and
          R. Smith\inst{4}
          \and
          G. Brunetti\inst{2}
          \and
          I. D. Roberts\inst{5}
          \and
          R. J. van Weeren\inst{5}
          \and
          K. Rajpurohit\inst{6}
          }
          
   \institute{
  INAF-Padova Astronomical Observatory, Vicolo dell’Osservatorio 5, I-35122 Padova, Italy\\
  \email{ alessandro.ignesti@inaf.it} 
\and 
INAF, Istituto di Radioastronomia di Bologna, via Gobetti 101, 40129 Bologna, Italy
\and
Dipartimento di Fisica e Astronomia ``Galileo Galilei'', Universit\'a di Padova, vicolo dell'Osservatorio 3, IT-35122, Padova, Italy
\and
Universidad T\`ecnica Federico Santa Mar\`ia
Avda. Vicu\~na Mackenna 3939, Oficina A035, San Joaqu\`in,
Santiago, Chile
\and
Leiden Observatory, Leiden University, PO Box 9513, 2300 RA Leiden, The Netherlands
\and
Harvard-Smithsonian Center for Astrophysics, 60 Garden Street, Cambridge, MA 02138, USA.
}
\authorrunning{Ignesti et al.}

   \date{Received November 28, 1988; accepted January 17, 1989}

 
  \abstract
   {Wide-field radio continuum observations of galaxy clusters are revealing an increasing number of spiral galaxies hosting tens of kpc-long radio tails produced by the nonthermal interstellar medium being displaced by the ram pressure; }
   {We present a semi-empirical model for the multi-frequency radio continuum emission from ram pressure stripped tails based on the pure synchrotron cooling of a radio plasma moving along the stripping direction with a uniform velocity;}
   {We combine LOFAR and uGMRT observations at 144 and 400 MHz to study the flux density and spectral index profiles of the radio tails of 7 galaxies in Abell 2255, and use the model to reproduce the flux density and spectral index profiles, and infer the stripped radio plasma velocity.}
   {For 5 out of 7 galaxies we observe monotonic decrease in both flux density and spectral index up to $~30$ kpc from their stellar disk. Our model reproduces the observed trends with a radio plasma bulk projected velocity between 160 and 430 km s$^{-1}$. This result represents the first indirect measure of the stripped, nonthermal interstellar medium velocity. The observed spectral index trends indicate that the synchrotron cooling is faster than the adiabatic expansion losses, thus suggesting that the stripped radio plasma can survive for a few tens of Myr outside of the stellar disk. This provides a lower limit for the lifetime of the stripped ISM outside of the disk. As a proof of concept, we use the best-fit velocities to constrain the galaxies' 3D velocity in the cluster to be in the 300-1300 km s$^{-1}$. We estimate the ram pressure affecting these galaxies to be between 0.1 and 2.9 $\times10^{-11}$ erg cm$^{-3}$, and measure the inclination between their stellar disk and the ram pressure wind.}
   {}

   \keywords{radio continuum: galaxies -- galaxies: clusters: Abell 2255 -- method: observational               }

   \maketitle
%

\section{Introduction}
Galaxies in clusters can evolve from star-forming into passive systems \citep[e.g.,][]{Dressler1980,Blanton_2009,Fasano2000,Cortese_2021}. One of the main drivers of this `environmental processing' is the interaction between the galaxy interstellar medium (ISM) and the intracluster medium (ICM) filling the cluster volume. This physical interaction manifests as an external pressure exerted by the ICM, the so-called `ram pressure' $P_{\text{Ram}}$ that can overcome the stellar disk binding force and strip the ISM components from the galaxy \citep[][]{Gunn1972}. This ram pressure stripping (RPS) can be crucial for the galaxy, as the gas loss can effectively quench the star formation in the stellar disk \citep[][for a review]{Boselli_2022}, thus making it an important quenching pathway for satellite galaxies \citep[e.g.,][]{Vollmer_2001,Tonnesen_2007,Vulcani2020,Watts_2023}. RPS is a frequent phenomenon, as almost all galaxies in clusters undergo a RPS event visible at optical wavelengths in their life \citep[][]{Vulcani_2022}. The ram pressure action is not limited to displacing the ISM outside of the disk.
The impact of ram pressure can result in many effects, including compression of gas along the leading edge of the disk \citep[e.g.,][]{Rasmussen_2006,Poggianti_2019,Roberts_2022e}, disturbed galaxy morphologies and trailing tails of stripped gas \citep[e.g.,][]{Kenney2004,vanGorkom_2004,Fumagalli2014,Poggianti2017}, and condensation of star-forming knots in the tails \citep[][]{Kenney2014,Poggianti2019}. It can also temporarily enhance the global star formation \citep[e.g.,][]{Poggianti_2016,Vulcani2018,Roberts_2020} and trigger the activity of the central nuclei \citep[e.g.,][]{Poggianti2017b,Peluso_2021}. Indeed it has been observed that it can affect the microphysics of the ISM on small scales, for examples by stimulating the conversion from atomic to molecular hydrogen \citep[][]{Moretti_2020}, enhancing the cosmic rays' diffusivity \citep[][]{Farber_2022,Ignesti_2022d}, or inducing the mixing between ISM and ICM \citep[][]{Sun_2022,Franchetto_2021}. The most extreme examples of galaxies undergoing strong RPS are the so-called jellyfish galaxies \citep{Fumagalli2014, Smith2010, Ebeling2014,Poggianti2017}. In the optical/UV band, these objects show extraplanar, unilateral debris extending beyond their stellar disks, and striking tails of ionised gas. Jellyfish galaxies mostly reside in galaxy clusters and are a transitional phase between infalling star-forming spirals and quenched cluster galaxies, hence they provide a unique opportunity to understand the impact of gas removal processes on galaxy evolution. 


A number of jellyfish galaxies have been observed showing tails of radio continuum emission extending for tens of kpc from their stellar disk \citep[e.g.,][]{GavazziJaffe1987,Murphy2009,Vollmer_2013,Chen_2020, Muller_2021,Roberts_2021b,Roberts_2021c,Ignesti_2022d}. These radio tails develop, typically, steep-spectrum emission ($\alpha<-0.9$ at GHz frequencies) within few tens of kpc from the stellar disk \citep[e.g.,][]{Vollmer_2004,Chen_2020,Ignesti_2021,Muller_2021,Lal_2022,Venturi_2022}. Therefore, they are best observed below GHz  frequencies and, for this reason, they are being observed more and more frequently by the LOFAR Two-metre Sky Survey \citep[LoTSS][]{Shimwell_2017,Shimwell_2019,Shimwell_2022} , which provide high resolution ($6 ``$) and highly sensitive ($\sim100$ $\mu$Jy beam$^{-1}$) images of the Northern sky at 120-168 MHz. Currently, about a hundred spiral galaxies with RPS radio tails have been reported, and thanks to this large statistic, it is now possible to conduct statistical studies of the RPS tail development in clusters \citep[e.g.,][for a recent example]{Smith_2022}. These `radio tails' are produced by cosmic ray electrons (CRe), accelerated to energies of a few GeV by supernovae explosions in the stellar disk \citep[][]{Condon_1992}. The relativistic plasma, together with the ISM, is then stripped from the disk by the ram pressure. The CRe cool down by emitting radio waves via synchrotron radiation until the stripped clouds evaporate in the ICM. The stripped tail magnetic field, which is responsible for the CRe synchrotron losses, can be further amplified by the ICM magnetic draping \citep[][]{Dursi_2008,Pfrommer_2010, Ruszkowski_2014,Muller_2021}. This qualitative scenario is supported by multi-frequency studies that observed a spectral index steepening with the radial distance along these tails \citep[e.g.,][]{Vollmer_2004,Chen_2020,Muller_2021,Ignesti_2021,Roberts_2021c,Venturi_2022}. In this framework, the radio tail length would be mainly driven by two factors, that are the synchrotron cooling time  \citep[][]{Pacholczyk_1970}, and the radio plasma bulk velocity, $V$, along the stripping direction. An interesting implication of this scenario is that, for the radio plasma to cool down mainly via synchrotron emission, the stripped ISM clouds should be able to survive in the ICM for a timescale that is, at least, as long as the radiative time. Constraining the stripped ISM clouds lifetime outside of the stellar disk is valuable for investigating the origin of the extraplanar star formation of jellyfish galaxies, or investigate general astrophysical problems such as the evolution of cold gas clouds in a hot wind.

In this work we develop a semi-empirical model to reproduce the radio tail of RPS galaxies and we test it on a sample of galaxies in the galaxy cluster Abell 2255 imaged with deep observations at 144 and 400 MHz. The manuscript is organized as follows. In Section \ref{data_analysis} we present the sample and the data used for this analysis, and the model developed to reproduce them. The results are presented in Section \ref{results_sec} and discussed in Section \ref{discussion}, where we present also the caveats of our work. Throughout the paper, we adopt a $\Lambda$CDM cosmology with $\Omega_{\text{DM}}=0.7$, $\Omega_{\text{matter}}=0.3$, and $H_0=70$ km s$^{-1}$ Mpc$^{-1}$, which yields $1''=1.512$ kpc at
the cluster redshift \citep[$z=0.08012$,][]{Golovich_2019}. We describe the radio synchrotron spectrum as $S\propto\nu^\alpha$, where $S$ is the flux density, $\nu$ the frequency, and $\alpha$ is the spectral index.
\section{Data analysis}
\label{data_analysis}
\begin{figure*}[t!]
    \centering
    \includegraphics[width=\linewidth]{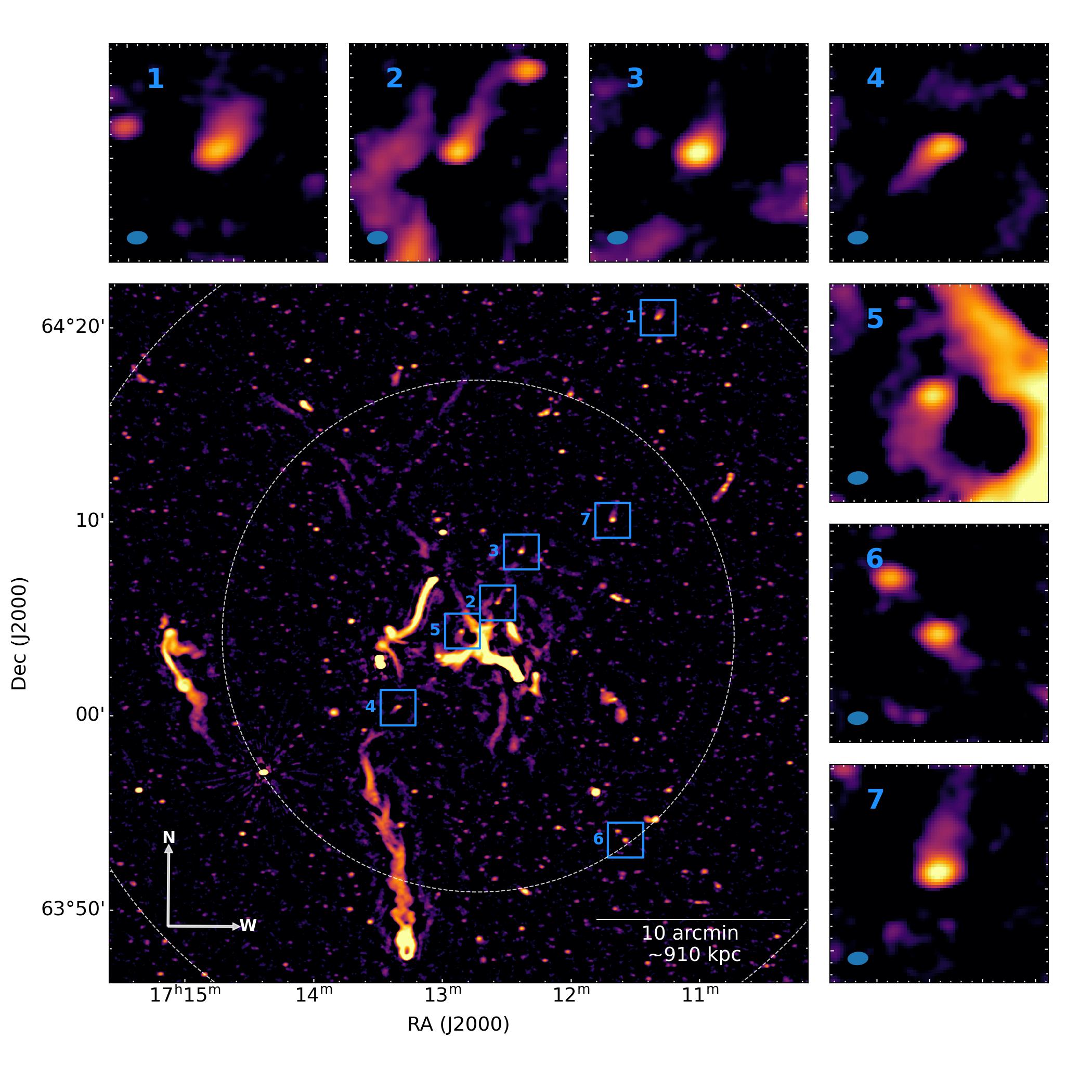}
    \caption{LOFAR image at 144 MHz of A2255 obtained with a lower uvcut of 2000 $\lambda$ (RMS=55 $\mu$Jy beam$^{-1}$, resolution $10.3``\times6.6``$). Inner and outer dashed circles point out $R_{500}=1.196$ Mpc and $R_{200}=2.033$ Mpc, respectively. The blue boxes point out the 7 galaxies analysed in this work, shown in the panels. The box size corresponds to the FOV of the corresponding panel.}
    \label{sample_img}
\end{figure*}
\subsection{Cluster properties and data preparation}

We analyze the radio emission of a sample of galaxies in the galaxy cluster Abell 2255 \citep[][hereafter A2255]{Abell_1958}, a nearby system \citep[$z=0.08012$,][]{Golovich_2019} with a complex merger dynamics. Optical analysis suggested a merger along the line of sight, whereas the X-ray morphology, elongated along the E-W axis, points to a second merger in that direction \citep[][]{Yuan_2003,Golovich_2019}. In the radio band, A2255 shows a diffuse radio halo at its center \cite[reported for the first time by ][]{Jaffe_1979}, and a large number of head-tail radio galaxies \citep[e.g.,][]{Harris_1980,Feretti_1997,Miller_2003,Pizzo_2009,Botteon_2020}. Thanks to deep observations provided by the LOw Frequency ARray \citep[LOFAR,][]{vanHaarlem_2013}, it has been revealed that A2255 hosts diffuse radio emission extended beyond $R_{200}$ \citep[][]{Botteon_2022}. We summarize in Table \ref{a2255} the properties of A2255.
\begin{table}[]
    \centering
    \begin{tabular}{cr}
    \toprule
         Property&Value  \\
    \midrule
        RA [$^\circ$]&258.216\\
        DEC [$^\circ$]&+64.063\\
        $z$ & 0.08012$\pm$0.00024\\
        $\sigma_{cl}$ [km s$^{-1}$]& 1137$\pm$50\\
        $M_{500}$ [$\times10^{14}$ $M_\odot$]&5.38$\pm$0.06\\
        $R_{500}$ [Mpc]& 1.196\\
        $R_{200}$ [Mpc]& 2.033\\
        
    \midrule
    \end{tabular}
    
    \caption{Summary of the properties of A2255. From top to bottom: equatorial coordinates \citep[][]{Eckert_2017}; redshift $z$ and velocity dispersion $\sigma_{cl}$ \citep[][]{Golovich_2019}; $M_{500}$, $R_{500}$ and $R_{200}$ \citep[][]{Planck_2016}; .  }
    \label{a2255}
\end{table}

This cluster is the ideal candidate to test our model due to the availability of deep LOFAR and upgraded Giant Metrewave Radio Telescope (uGMRT) observations at, respectively, 120-168 MHz and 400 MHz. 
We make use of the 75 hrs LOFAR observation presented in \citet[][for a detailed description of the data calibration]{Botteon_2022} to produce a new image at a central frequency of 144 MHz with a lower uvcut of 2000 $\lambda$, {\tt ROBUST}=-0.5 \citep[][]{Briggs_1994}, and a taper of 5 arcsec. The image is centered on the cluster and covers an area of 1 deg$^2$. The imaging was carried out with the software {\tt WSCLEAN} \citep[][]{Offringa_2014}. The resulting image, shown in Figure \ref{sample_img}, has an angular resolution of $10.3''\times6.6''$, and RMS=55 $\mu$Jy beam $^{-1}$. The lower uvcut permits us to remove the radio emission diffused on scales larger than $\sim100''$ ($\sim150$ kpc) and, thus, to reveal the smaller sources beneath, such as the radio tails. 

A2255 was observed with the uGMRT in band 3 (300--500 MHz) for 40h (project code: 39\_032, PI: A. Botteon), and presented here for the first time. The observations were divided into 4 runs of 10~h each, carried out on 2021 February 20, 26, March 13, and April 01, bookended by two 8~m scans on the flux density calibrators 3C286 and 3C48. The data were recorded in 2048 frequency channels with an integration time of 5.3~s using both the narrow-band (bandwidth of 33.3 MHz) and wide-band (bandwidth of 200 MHz) backends. The data were processed with the Source Peeling and Atmospheric Modeling (SPAM) package \citep[][]{Intema_2009}, which is a widely used software to reduce GMRT observations that performs calibration of the flux density scale, correction for the bandpass, data averaging and flagging, and direction-dependent calibration. In order to produce deep and wide-band images of A2255 at the central frequency of 400 MHz, we proceed as follows. First, we processed the 4 narrow-band datasets independently to assess the quality of the 4 observing runs. Second, we split each wide-band dataset into 6 slices with equal bandwidth of 33.3 MHz, from 300 to 500 MHz. This step is necessary for SPAM to handle the wide-band of the new uGMRT backend and it has already been adopted in previous studies \citep[e.g.,][]{Botteon_2020b, DiGennaro_2021,Schellenberger_2022}. Third, we obtained a global sky model from the best narrow-band image (run of 2021 April 01) to jointly calibrate and merge the slices centered at the same frequency of the 4 observing runs. Fourth, we jointly deconvolved the 6 resulting calibrated slices with {\tt WSClean} enabling the multiscale multifrequency deconvolution \citep[][]{Offringa_2017}. The resulting image, which was corrected for the primary beam response, has a resolution of $9.8''\times9.0''$ and and RMS=20 $\mu$Jy beam $^{-1}$. In this paper we focus on the jellyfish galaxies, while a forthcoming publication will be focused on the diffuse radio emission (Rajpurohit et al., in prep)

We also make use of images and spectroscopy from the 12$^{th}$ release of the Sloan Digital Sky Survey \citep[SDSS,][]{Donald_2000}. We combine SDSS images\footnote{From \url{https://dr12.sdss.org/mosaics/}} in the {\tt g-}, {\tt r-} and {\tt i}-filter to produce color images of the cluster galaxies.
\subsection{Galaxy sample selection}
\label{sample_intro}
We select a sample of suitable galaxies for this analysis solely on the basis of their radio emission morphology. Specifically, we focus on those galaxies with an evident spiral/disk morphology in the SDSS image, and radio tails that are detected above the 3$\times$RMS level and resolved by more than 3 resolution elements at 144 MHz. This latter condition is necessary to reliably sample their flux density profiles. 

We end up with a sample of 7 galaxies which we show in Figure \ref{sample_img}. Their properties are summarised in Table \ref{sample_tab}. According to SDSS 16 classification \citep[][]{SDSS16}, these galaxies are star-forming spirals without evidence of AGN activity. Hence, their radio tails should be produced solely by the interactions between the ram pressure winds and the CRe accelerated by the supernovae. All the galaxies have the radio tail pointing away from  the cluster center, and, with the only exception of $\#6$, they are remarkably aligned along the cluster NW-SE axis, which may suggest that they are collectively falling toward A2255 along a privileged direction. This piece of evidence might indicate that the NW-SE axis is critical for the ongoing merger, thus adding another piece of the puzzle to the complex dynamics of this cluster. Finally, we note that they are all blue-shifted with respect of the cluster $(z<z_{cl})$.
\begin{table*}[]
    \centering
    \begin{tabular}{ccccc}
    \toprule
    ID&Name & RA, DEC & $z$ & $R_{CL}$ \\
    && [$^\circ$, $^\circ$]&&[kpc]\\
    \midrule
1&LEDA 2667121&257.823, 64.342&0.0768&1781.66\\
2&LEDA 3138983&258.141, 64.098&0.0762&259.13\\
3&[PVK2003] 258.09561 +64.14133&258.095, 64.141&0.0752&514.6\\
4&[YZJ2003] 2- 63&258.334, 64.009&0.0799&406.32\\
5&LEDA 59848 &258.213, 64.073&0.0738&55.93\\
6&[YZJ2003] 2-130&257.894, 63.894&0.0767&1198.62\\
7&LEDA 2665175&257.914, 64.169&0.0768&920.2\\
     \midrule
    \end{tabular}
    \caption{Sample of galaxies studied in this work. From left to right: identification number used in this paper; Galaxy ID; Coordinates of the first sampling bin; SDSS redshift; Projected clustercentric distance.}
    \label{sample_tab}
\end{table*}

\subsection{Analysis of the radio tails}
\subsubsection{Flux density and spectral index profiles}
In order to evaluate the properties of the radio tails, we sample their radio emission to infer the flux density decline with the distance. We define a grid composed of a set of aligned elliptical regions with a size of $11\times7$ arcsec$^2$, so larger than the radio images' beam sizes, that start from the galaxy center and  extend to the end of the tail as observed at 144 MHz. The first bin, that is the one located over the stellar disk, serves solely to define the galaxy position  in the cluster to compute the projected length of the tail. For each radio map, we proceed measuring the flux density in each elliptical region, beyond the first one, located over the corresponding 3$\sigma$ contour, and we compute the corresponding uncertainty $\sigma=\sqrt{A/A_{\text{beam}}}\times$RMS (where $A$ and $A_{\text{beam}}$ are, respectively, the area of the sampling bin and the beam). Then we measure the spectral index in each bin in which we observe emission both at 144 and 400 MHz as:
\begin{equation}
\alpha=\frac{\log\left(\frac{S_{144}}{S_{400}}\right)}{\log\left(\frac{144}{400}\right)}\pm\frac{1}{\log\left(\frac{144}{400}\right)}\sqrt{\left(\frac{\sigma_{144}}{S_{144}}\right)^2+\left(\frac{\sigma_{400}}{S_{400}}\right)^2}\text{,}
    \label{alfa}
\end{equation}
where $\sigma_{144}$ and $\sigma_{400}$ are the flux density uncertainties at 144 and 400 MHz. Finally, in order to define the distance of each bin from the stellar disk, we consider the projected physical distance between the center of the each bin and that of the previous one. This approach permits us to compute the position of each bin with respect to the stellar disk as the sum of the distances of the previous bins. 

The radio tail projected length lies between 35 and 60 kpc, with galaxies showing different declining slopes. For galaxy $\#2$  we adopt special considerations because the radio tail connects with another galaxy located toward the north-west (Figure \ref{sample_img}). The current image resolution does not allow us to discriminate if the connection is real or just a blend of multiple sources. 
Regardless, we decide to sample its tail up to $\sim40$ kpc from the disk. 
\subsubsection{Modeling the multi-frequency radio tail}
\label{model}
We model the expected flux density and spectral index profiles by assuming that they are produced by radio plasma clouds, accelerated in the stellar disk, that move along the stripping direction with a uniform bulk velocity and that the CRe cooling is dominated by synchrotron and Inverse Compton energy losses due to the Cosmic microwave background (CMB) \citep[e.g.,][]{Longair_2011}. This  assumption is motivated by the fact that the stripped tails are mostly composed of ionized gas \citep[e.g.,][]{Boselli_2022}, thus the CRe perceive a local gas density that is lower than within the stellar disk ($n_{\text{gas}}\sim10^{-2}-10^{-1}$ cm$^{-3}$). Under such conditions, the time scales of the other energy loss mechanisms, such as ionization losses, bremmstrahlung radiation, or Inverse Compton with the galactic radiation field, are of the order of a few Gyr \citep[e.g.,][Equation 5-8]{Basu_2015}, and therefore they are negligible with respect to synchrotron losses ($t\simeq10^7-10^8$ yr). Our estimate is conservative because we assume a uniform magnetic field and we neglect adiabatic losses. The latter assumption is partially supported by the fact that multi-frequency studies of ram pressure stripped galaxies revealed a spectral index decline with the distance \citep[][]{Vollmer_2013,Muller_2021,Roberts_2021c,Ignesti_2022b}, thus suggesting that the synchrotron losses time-scale is shorter than the one of the adiabatic losses. 
Under these conditions, the synchrotron emissivity spectrum $j(\nu/\nu_{\text{br}})$ is defined as:
\begin{equation}
    j\left(\frac{\nu}{\nu_{\text{br}}}\right)=\sqrt{3}\frac{e^3}{m_ec^2}\int_{0}^{\frac{\pi}{2}}\text{sin}^2\theta\text{d}\theta\int_0^{+\infty} n(\gamma)F\left(\frac{\nu}{\nu_{\text{br}}}\right)\text{d}\gamma\text{ ,}
    \label{eq}
\end{equation}
where $\theta$ is the pitch angle, $\nu_\text{br}=\frac{3}{2}\frac{eB\sin \theta}{m_e c}\gamma^2$ is the break frequency, $F(\nu/\nu_\text{br})$ is the synchrotron kernel function \citep[][]{Rybicki-Lightman_1979}, $e$ and $m_e$ are the electron charge and mass, respectively, and $n(\gamma)\propto \gamma^\delta$ is the CRe energy distribution. We compute a sampled spectrum for $j(\nu/\nu_{\text{br}})$ by solving numerically Equation \ref{eq} under the assumptions of $\delta=-2.2$, which entails an injection index $\alpha=(\delta+1)/2=-0.6$, and the favorable minimal energy loss magnetic field $B=B_\text{CMB}/\sqrt{3}\simeq2.2$ $\mu$G, where $B_{\text{CMB}}=3.25(1+z)^2\simeq3.8$ $\mu$G is the CMB equivalent magnetic field. The latter entails that we are assuming the maximum CRe radiative time. The implications of our assumptions are discussed in Section \ref{caveats}.

\begin{figure*}
    \centering
    \includegraphics[width=.9\linewidth]{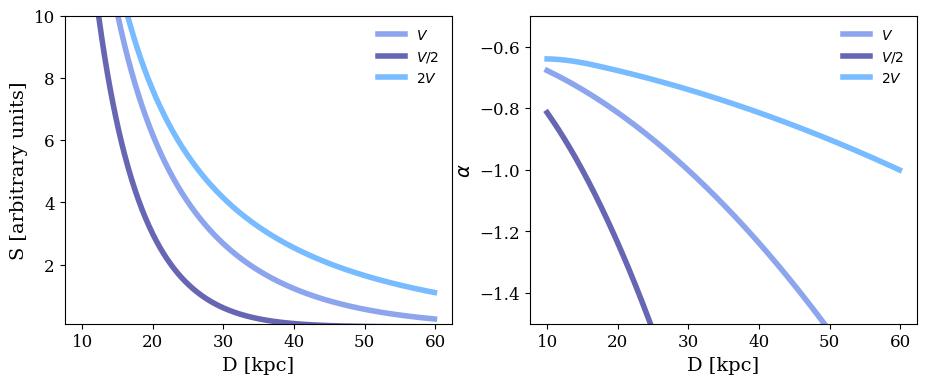}
    \caption{Flux density (left) and spectral index (right) model profiles for different values of the  velocity $V$ given a magnetic field $B_{\text{min}}$.}
    \label{spectrum}
\end{figure*}

In order to associate the emissivity spectrum to the observed flux density profiles, it is necessary to assume a `bulk velocity' for the radio plasma along the stripping direction. The velocity defines the radio plasma `dynamic age', that is the time elapsed since  it left the stellar  disk. For simplicity, we assume that the radio plasma moves with a uniform velocity $V$ along the stripping direction, and that it leaves the stellar disk immediately after being injected in the ISM. Consequently, the time elapsed since the CRe injection can be estimated as $\tau= D/V$, where $D$ is the (projected) distance from the stellar disk, and $V$ is the (projected) CRe bulk velocity with respect to the galaxy. We note that both $D$ and $V$ are projected quantities, but their ratio is equivalent to the ratio of the deprojected values. Therefore, the observed projected distances can be associated with a corresponding $\tau$. Then, to derive the corresponding model emissivity we make use of the radiative time definition: 
\begin{equation}
t_{\text{rad}}\simeq3.2\times 10^{10}\frac{B^{1/2}}{B^2+B_{\text{CMB}}^2} \frac{1}{\sqrt{\nu_{\text{br}} (1+z)}}\text{ yr,}
\label{cool_RW}
\end{equation}
where the magnetic fields are expressed in $\mu$G and the observed frequency $\nu_{\text{br}}$ is in units of MHz \citep[][]{Miley_1980} to compute the corresponding $t_\text{rad}$. Under the assumption that the radiative age of the plasma coincides with the time elapsed since the injection, i.e. $t_\text{rad}\simeq \tau=D/V$, for a given $V$ Equation \ref{cool_RW} associates to each spatial bin a corresponding $\nu_{\text{br}}(D)$ that, in turn, defines a value of $\nu/\nu_{\text{br}}(D)$ for a given $\nu$. Finally we compute the emissivity by interpolating the corresponding $j(\nu/\nu_{\text{br}}(D))$ from the sampled emissivity spectrum derived from Equation \ref{eq}. We also introduce a normalization factor $A$ that incorporates the conversion from emissivity to observed flux density. This procedure provides us with a flux density model profile as function of the distance that we can use to fit the observed flux density profiles to constrain $V$. By combining the interpolated emissivity values for two different frequencies at each distance, Equation \ref{alfa} permit us to model also the expected spectral index profile $\alpha(D)$ which we used to fit the observed spectral indexes. In general, this model predicts that flux density decreases monotonically with the distance, with a consequent spectral index steepening. The trend is tuned by the velocity, where the lower (higher) the value of $V$, the steeper (flatter) are the resulting profiles. As a consequence of our assumption $\delta=-2.2$, the flattest spectral index value allowed by this model is $\alpha=-0.6$. For reference, we show in Figure \ref{spectrum} how the resulting flux density and spectral index profiles change for different values of $V$ for a fixed $B=B_{\text{min}}$.

We use a least-square fit to constrain the two free parameters, $V$ and $A$, that adapt our sampled model profiles to the observed ones. Basically, the first one tunes the model steepening along the x-axis, whereas the second one matches model and observations along the y-axis. For each galaxy, we fit the 144, 400 MHz flux density, and spectral index profiles independently to see whether the model can infer consistent velocities. We take into account both the uncertainties for the flux density and spectral index to compute the error on the velocity and the 1$\sigma$ confidence interval for the best-fitting model.

\section{Results}
\begin{table}
    \centering
    \begin{tabular}{cccc}
    \toprule
    ID& $V_{144}$ & $V_{400}$ &$V_\alpha$ \\
    \midrule
1&667$\pm$316&429$\pm$83&206$\pm$33\\
2&654$\pm$407&3519&295$\pm$93\\ 
3&149$\pm$16&215$\pm$32&136$\pm$15\\ 
4&307$\pm$65&325$\pm$38&718\\
5&3087&174$\pm$28&718\\
6&304$\pm$132&160$\pm$32&993\\ 
7&402$\pm$104&251$\pm$44&172$\pm$26\\ 
\midrule
    \end{tabular}
    \caption{Best-fitting velocities in units of km s$^{-1}$ derived from the different profiles and the 1$\sigma$ uncertainties for the systems with sufficient statistics. }
    \label{velocisit}
\end{table}

\label{results_sec}

\begin{figure*}
\begin{multicols}{2}
\centering
\includegraphics[width=.9\linewidth]{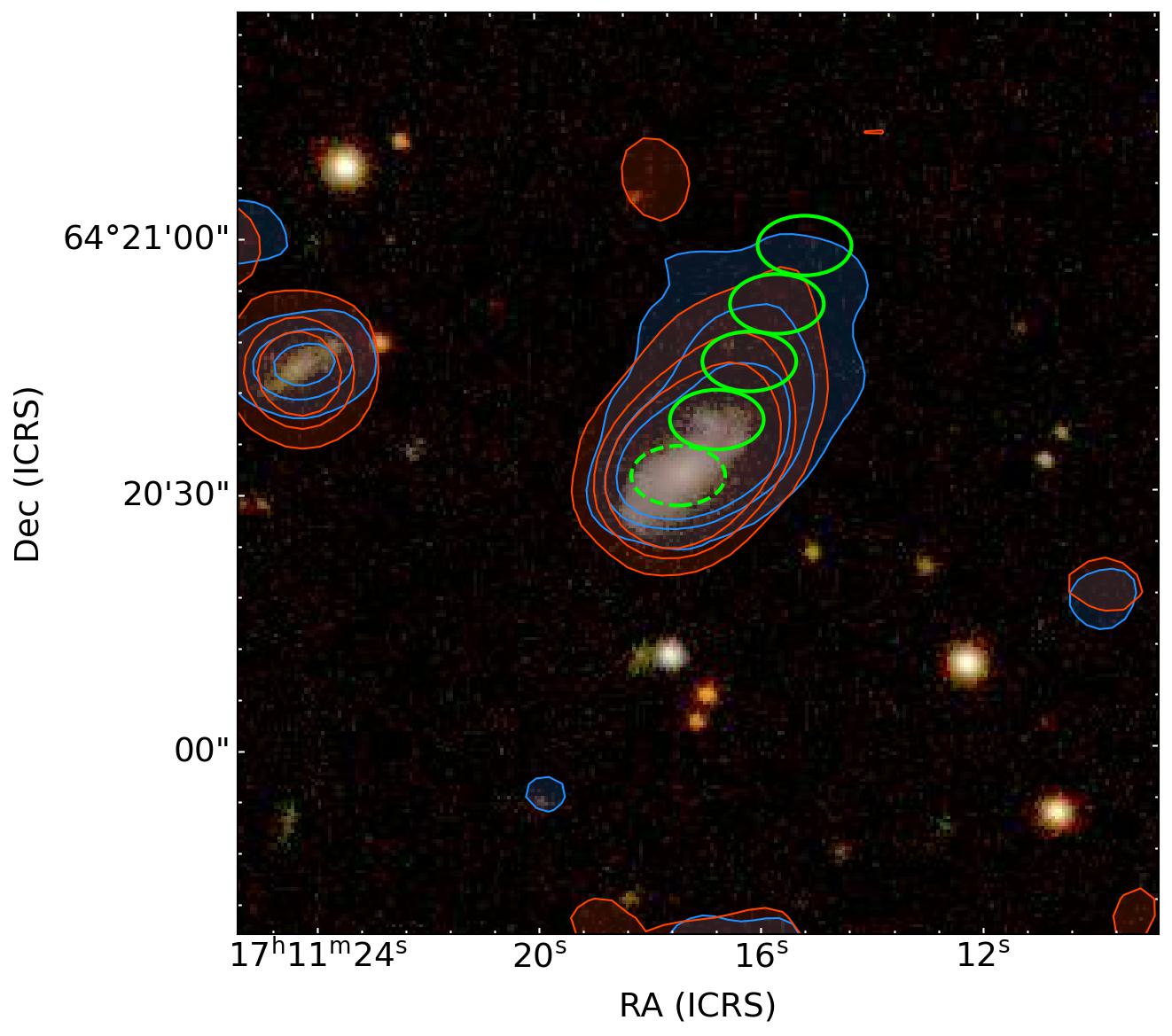}\par
\centering
\includegraphics[width=.9\linewidth]{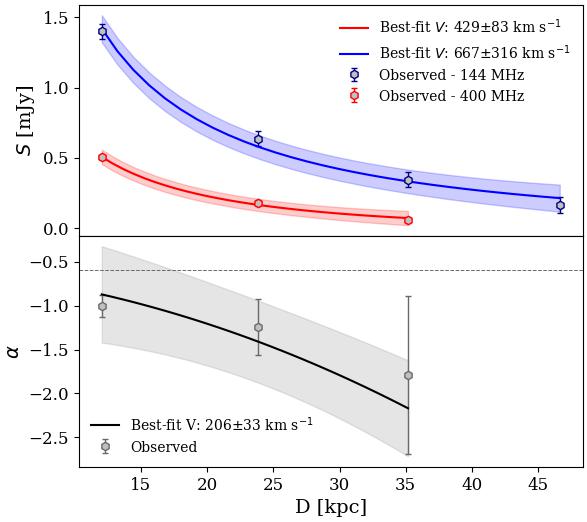}
\end{multicols}
\begin{multicols}{2}
\centering
\includegraphics[width=.9\linewidth]{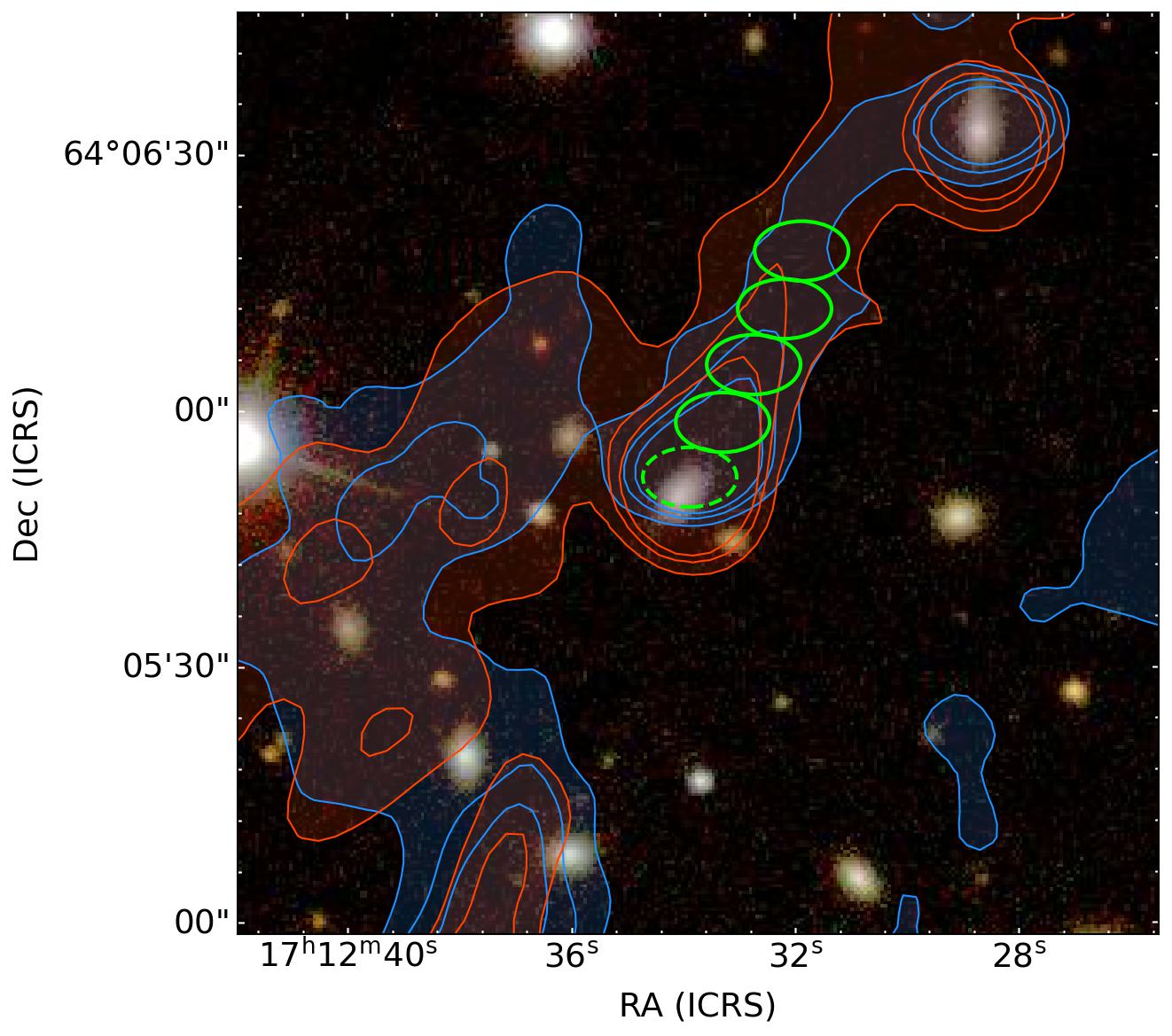}\par
\centering
\includegraphics[width=.9\linewidth]{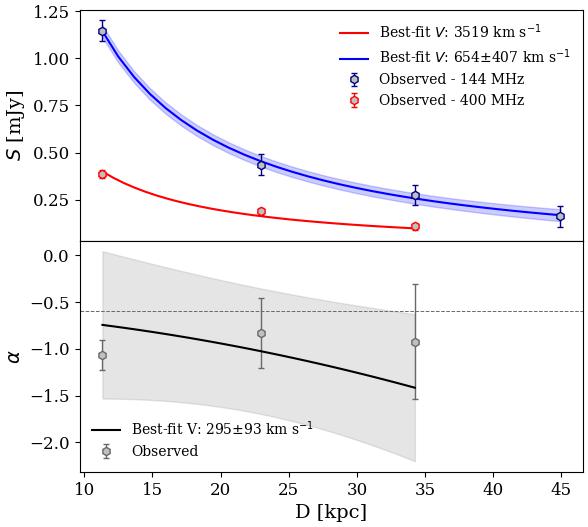}
\end{multicols}
\begin{multicols}{2}
\centering
\includegraphics[width=.9\linewidth]{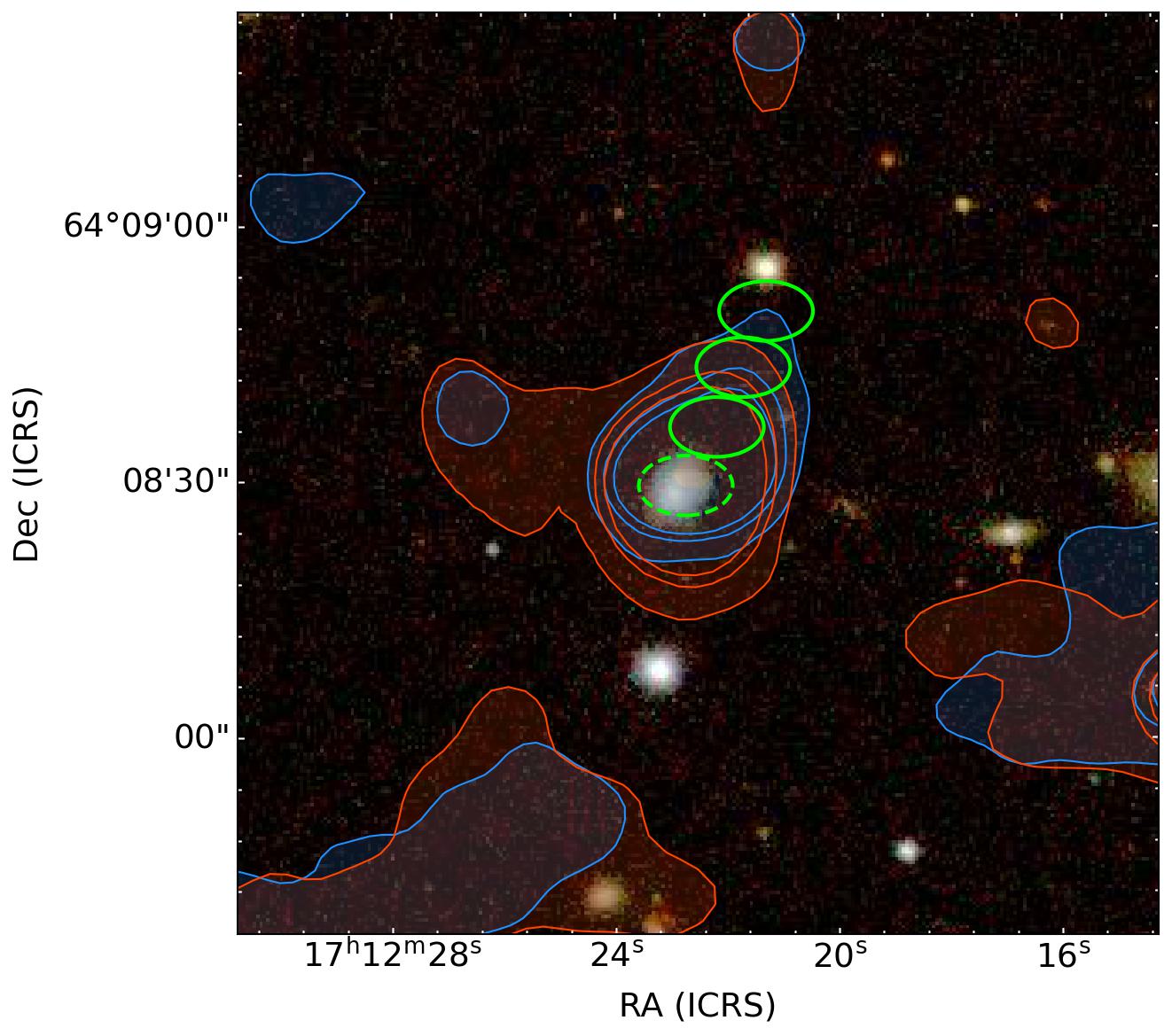}\par
\centering
\includegraphics[width=.9\linewidth]{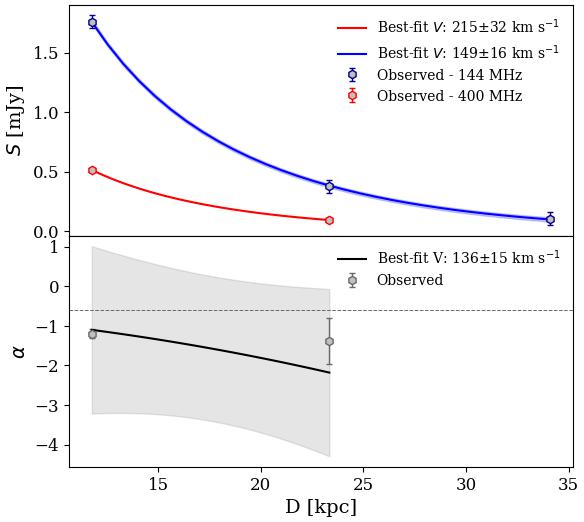}
\end{multicols}
\caption{From top to bottom: velocity fit for galaxy $\#1$, $\#2$, and $\#3$. Left: SDSS RGB image overlapped with the 3, 9, 15$\times$RMS surface brightness contours at 144 (blue) and 400 MHz (orange) and the sampling regions (green ellipses). The dashed region marks the reference center of the galaxy; Right: Flux density (top) and spectral index (bottom) trends with the distance from the stellar disk, and the corresponding best-fit profiles. The color-filled area indicates the $1\sigma$ confidence region. The horizontal dashed line points $\alpha=-0.6$.}
\label{best-fit}
\end{figure*}
\renewcommand{\thefigure}{\arabic{figure} (continued)}
\addtocounter{figure}{-1}

\begin{figure*}[tp]
\begin{multicols}{2}
\centering
\includegraphics[width=.9\linewidth]{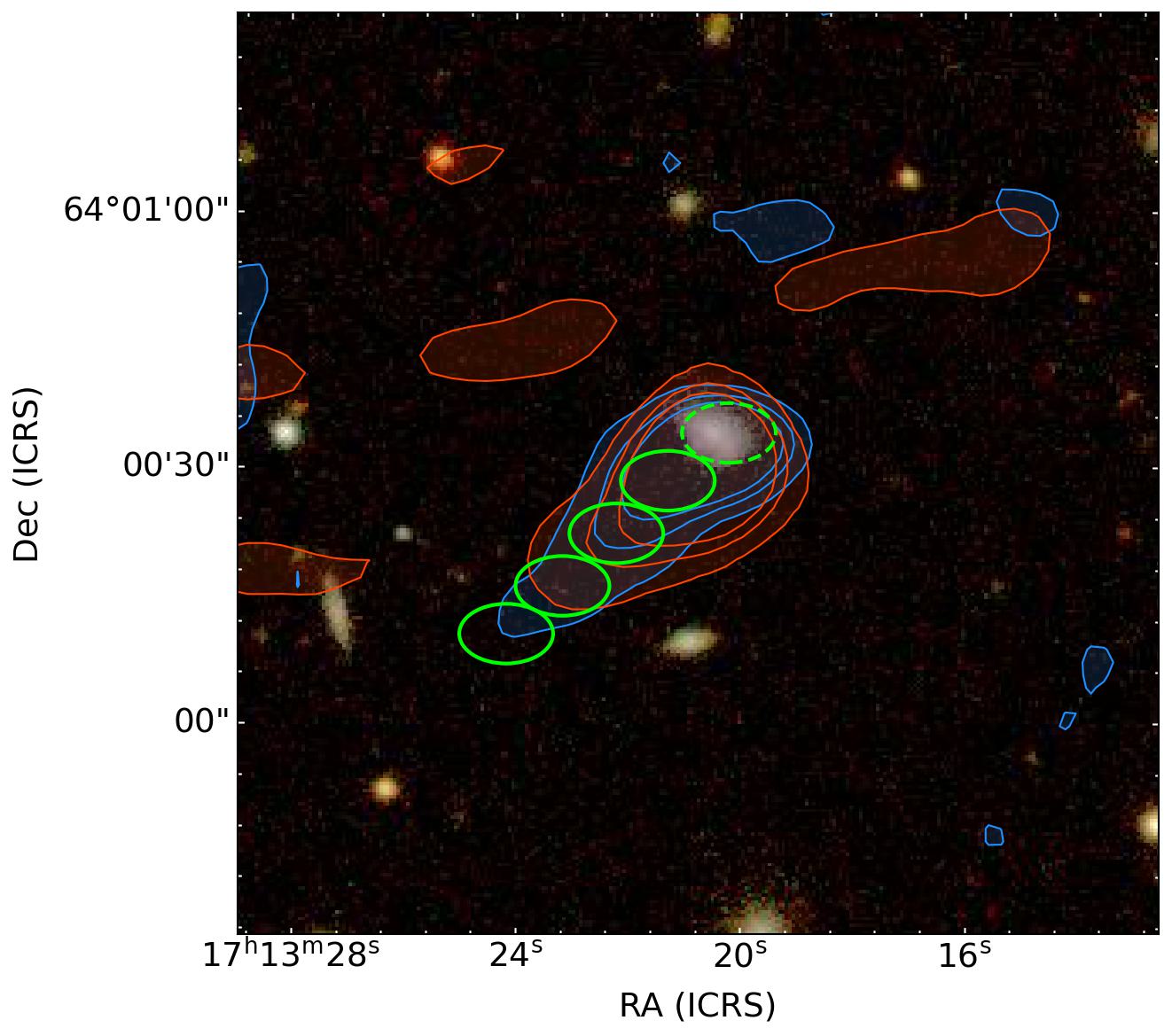}\par
\centering
\includegraphics[width=.9\linewidth]{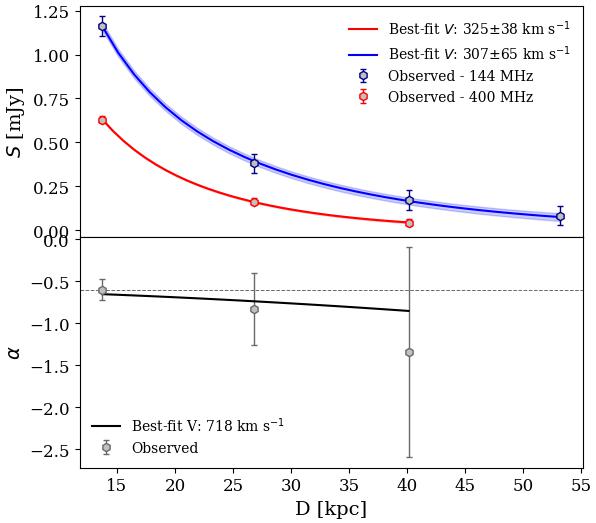}
\end{multicols}
\begin{multicols}{2}
\centering
\includegraphics[width=.9\linewidth]{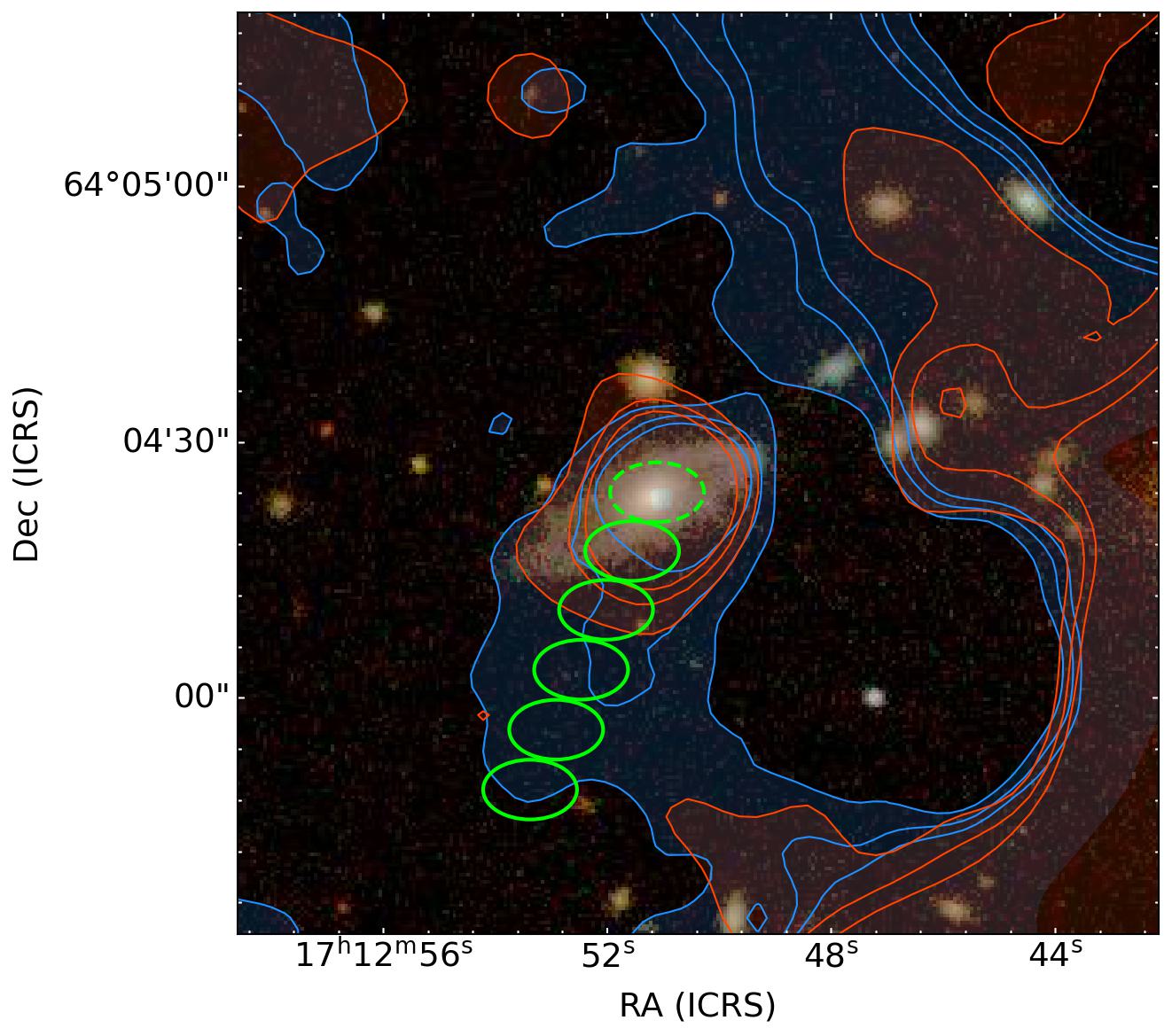}\par
\centering
\includegraphics[width=.9\linewidth]{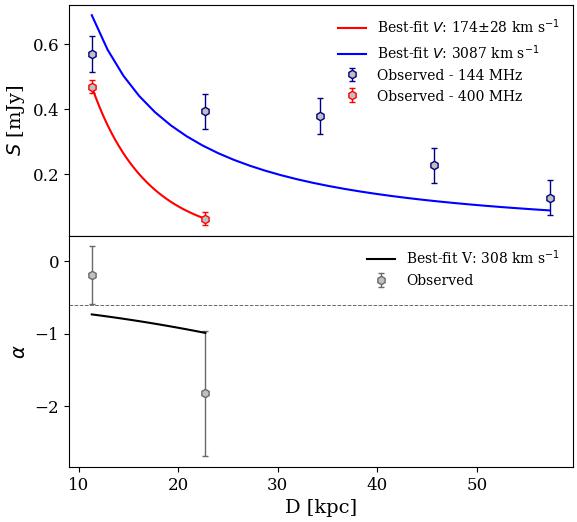}
\end{multicols}
\begin{multicols}{2}
\centering
\includegraphics[width=.9\linewidth]{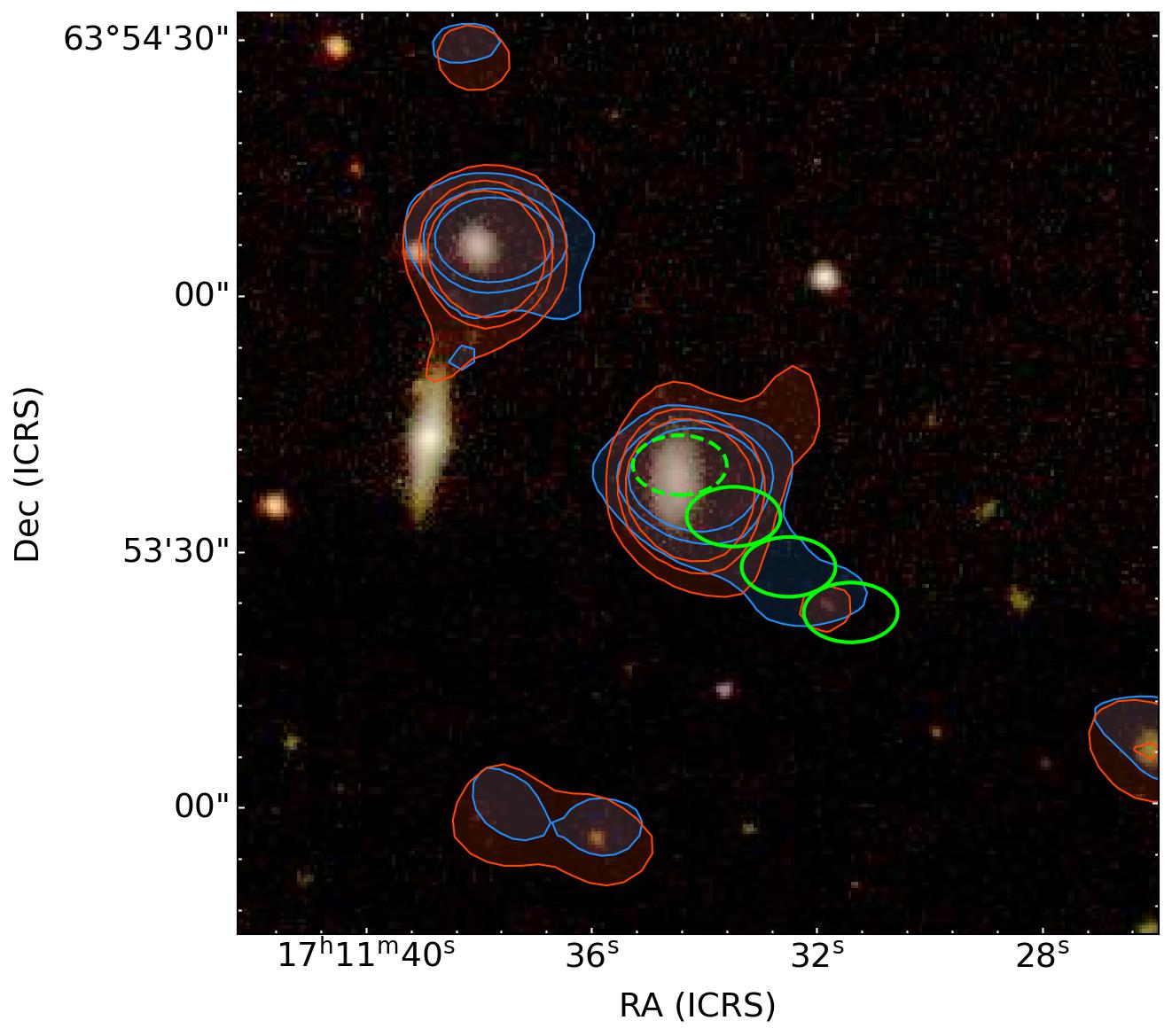}\par
\centering
\includegraphics[width=.9\linewidth]{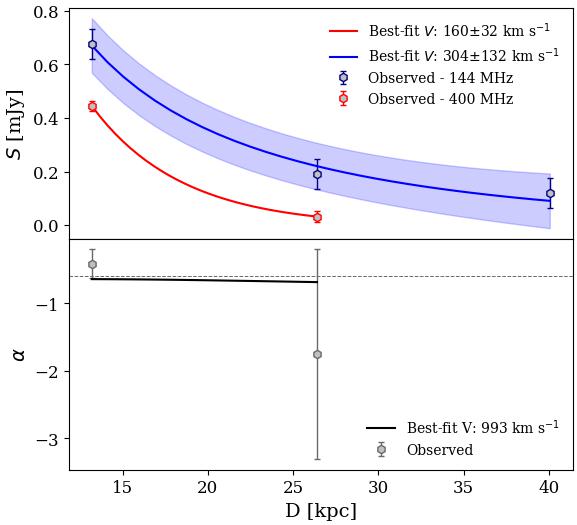}
\end{multicols}
\caption{From top to bottom: velocity fit for galaxy $\#4$, $\#5$, and $\#6$.}
\end{figure*}
\addtocounter{figure}{-1}
\begin{figure*}[tp]
\begin{multicols}{2}
\centering
\includegraphics[width=.9\linewidth]{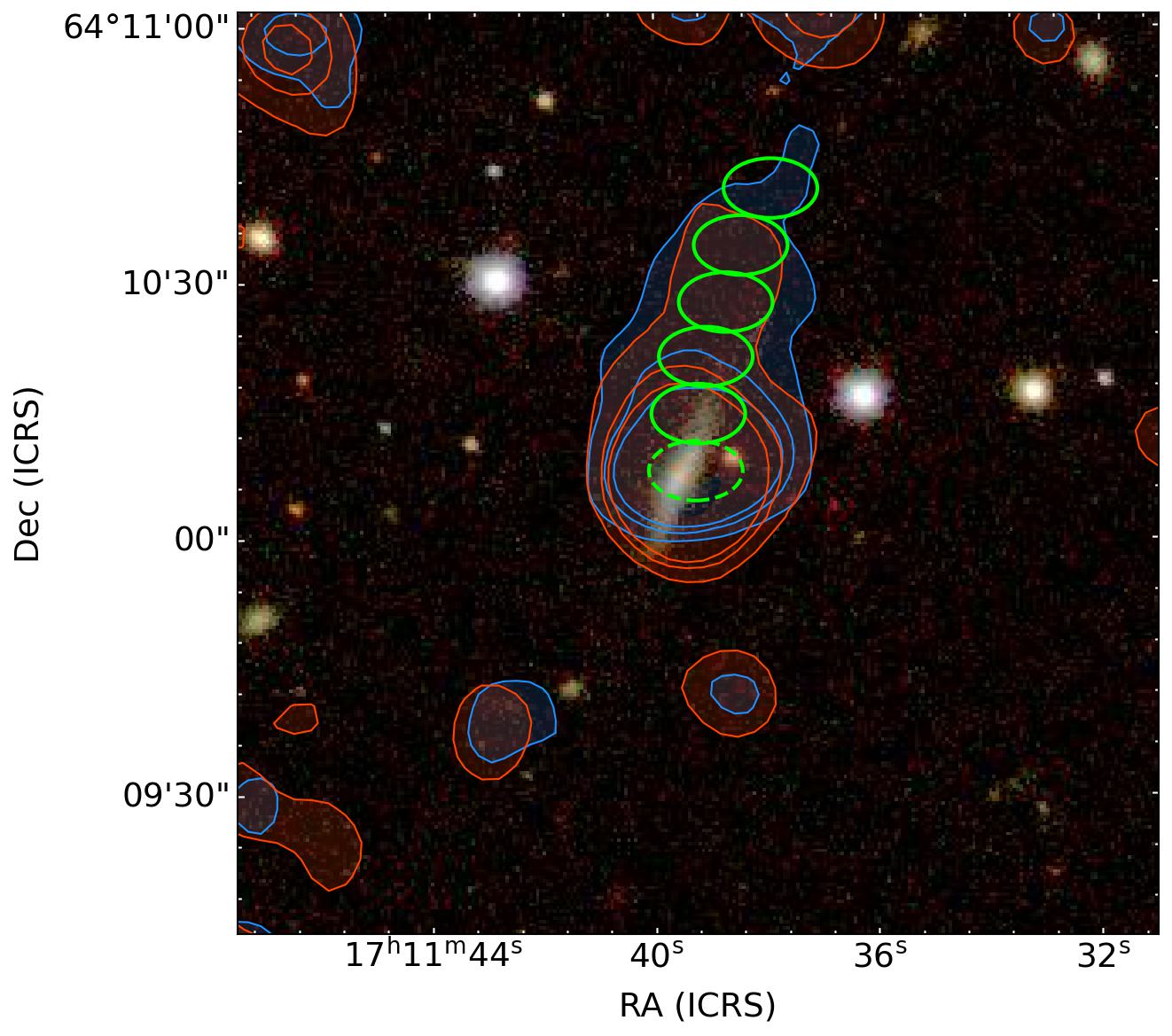}\par
\centering
\includegraphics[width=.9\linewidth]{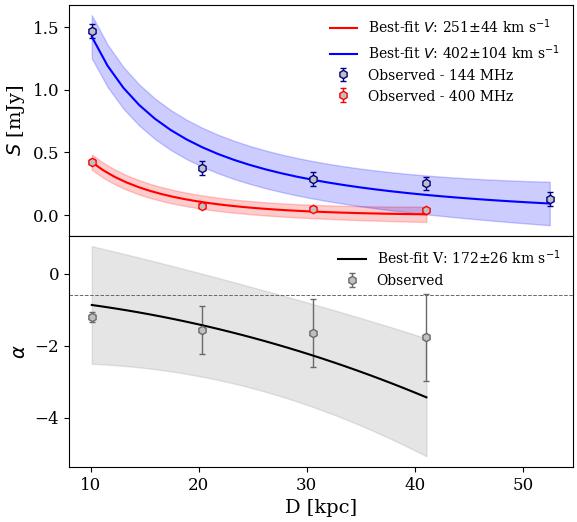}
\end{multicols}
\caption{Velocity fit for galaxy $\#7$.}
\end{figure*}
\renewcommand{\thefigure}{\arabic{figure}}
We report in Figure \ref{best-fit}, for each galaxy, the sampling grid over-imposed on the SDSS image and the 144 MHz (blue) and 400 MHz (red) contours, the corresponding flux density and spectral index profiles, and the best fit with the computed values of $V$. The results are summarized in Table \ref{velocisit} and Figure \ref{result_fig}. Due to the low number of bins of some galaxies, the best-fit uncertainties cannot be computed for every system. 

Concerning the observed profiles, we note that, in general, every galaxy shows a decreasing flux density profile at both frequencies. The 400 MHz profiles are systematically shorter than the 144 MHz one, which is consistent with the fact that the CRe emitting at higher frequency, in a uniform magnetic field, have shorter radiative times of those emitting at lower frequencies (Equation \ref{cool_RW}), and therefore they can travel for shorter distances. The spectral index profiles seem to steepen with the distance in 4 out of 7 galaxies.

About the model fitting, we observe that, albeit the 144 MHz profiles have a higher number of bins than the 400 MHz one and, thus, they should result in more solid results, the 400 MHz fit tend to have smaller uncertainties. We conclude that this is due to the fact that the curvature of the 400 MHz profiles is more evident of that of the 144 MHz ones, thus easing the fit convergence. The outcomes of the model fitting are varied:
\begin{itemize}
\item For galaxies $\#1$, $\#3$, and $\#7$ the velocities derived from the three tracers are consistent within 2$\sigma$;
\item For galaxies $\#4$ and $\#6$ the results are less clear. On the one hand, the flux density fit produces consistent results for the 144 and 400 MHz profiles. On the other hand, the spectral index fitting does not converge due to the fact that the values in the first bins are flatter than -0.6, which is the upper limit permitted by our model. Observing such a flat value of spectral index suggests that, inside those bins, the radio emission at 144 MHz either has an injection index flatter than -0.6, or it has been affected by ionization losses, which are not included in our simple model. Nevertheless, the spectral index profiles steepen with the distance, that is in agreement with our predictions;
\item For galaxies $\#2$ and $\#5$  results are inconclusive because either the flux density or the spectral index profiles are not decreasing monotonically. In both cases, we conclude that this is due to the fact that the radio tails are not real and/or induced by the RPS. For galaxy $\#2$, we suggest that the putative radio tail is instead the result from the emission from two galaxies that is blended due to the insufficient resolution of our images. For galaxy $\#5$, we conclude that the `tail' is actually mostly composed of  emission coming from the near, bright radio galaxy. Consequently, the fit could not converge and it returns nonphysical and not consistent values of $V$. Therefore, we exclude these two galaxies from the following Discussion. 
\end{itemize}
\begin{figure*}
\centering
   \includegraphics[width=.82\linewidth]{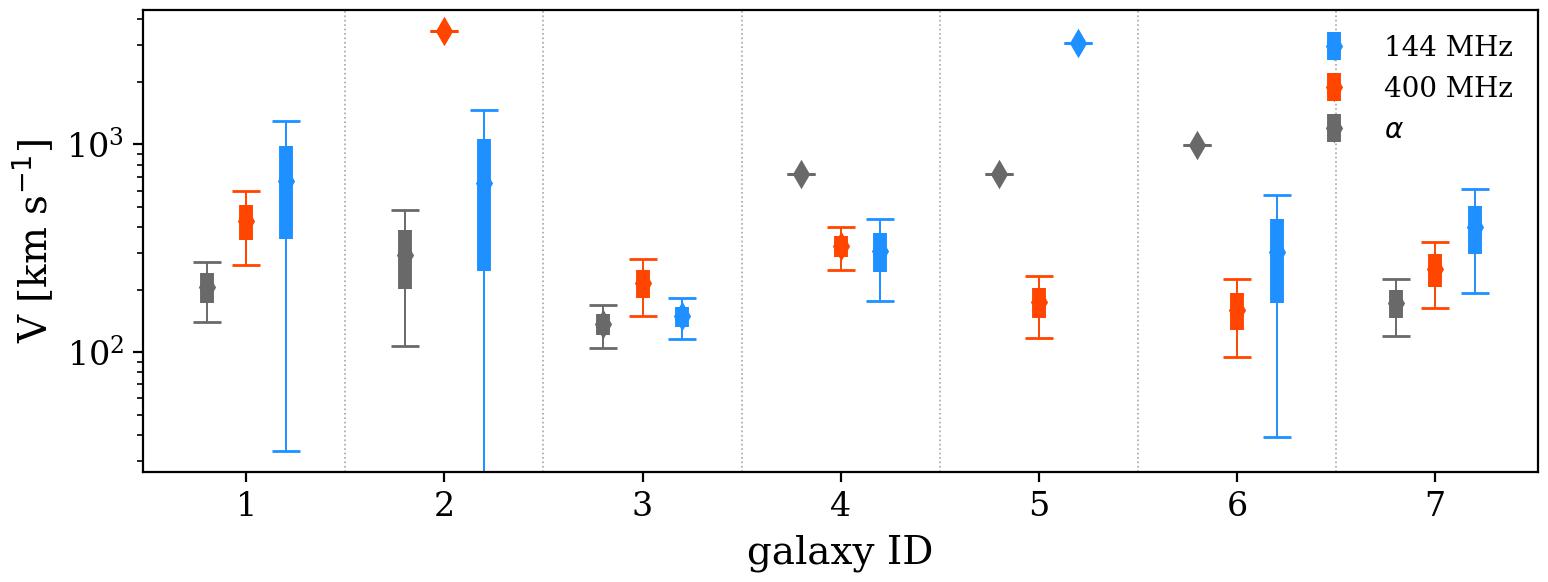}
    \caption{Best-fitting $V$ for each galaxy measured from the 144 (blue), 400 MHz (red) and spectral index (grey) profiles. The thick and thin error bars represent, respectively, the 1 and 2$\sigma$ confidence intervals. }
    \label{result_fig}
\end{figure*}
\section{Discussion}
\label{discussion}
\subsection{Insights into the properties of the stripped tails}
\label{disc_tail}
\subsubsection{The lifecyle of the radio plasma}
The semi-empirical model introduced in Section \ref{model} reproduces the radio tails' profiles of 5 out of 7 radio tails in A2255. This result permits us to investigate the physical properties of the radio-emitting CRe and the stripped ISM. To begin with, the best-fit velocities are of the order of the hundreds of km s$^{-1}$ (see Table \ref{velocisit}). These values support the previous results that constrained the CRe bulk velocity to be of the order of the ram pressure winds \citep[][]{Ignesti_2022b}. Our semi-empirical model provides an accurate measure of the CRe bulk velocity in these extraplanar fields that can be used to constrain numerical simulations of the nonthermal ISM components subjected to ram pressure \citep[e.g.,][]{Tonnesen_2014,Muller_2021,Farber_2022}. 

Concerning the radio plasma properties, in general we observe that the flux density and spectral index profiles decline monotonically with the distance, in agreement with the simple, pure cooling model. The emerging picture is that the radio plasma stripped from the stellar disk can cool down within the first tens of kpc via synchrotron emission before being dissipated by  adiabatic expansion or  mixing with the ICM. This confirms that the adiabatic losses timescale $t_{\text{ad}}$ is larger than the CRe radiative time $t_{\text{rad}}$ at 144 MHz, thus it tentatively constrains $t_{\text{ad}}>100$ Myr (Equation \ref{cool_RW}). 

The observed profiles can constrain the action of `re-energization' processes that would induce deviations from the monotonic decline. These include star formation taking place outside of the stellar disk, and  (re-)acceleration processes induced by shocks and turbulence within the tail. The first one is expected to interfere with the cooling by injecting fresh electrons along the tail, thus forming a `bump' in the flux density profile and a flattening in the spectral index one \citep[e.g., the case of JO206, see][]{Muller_2021}. In our sample, only galaxy $\#7$ shows a potential signature of this process for $D>30$ kpc, beyond which the observations slightly diverge from the model. This feature could be signature of either presence of `fresh' CRe produced in-situ by extraplanar star formation, or that the CRe losses may be dominated by adiabatic losses instead of synchrotron/IC (e.g., due to increasingly longer radiative times as consequence of a decrease of the magnetic field with the distance). However, we note that not observing a clear bump for the other galaxies does not rule out the possibility of having star formation in the stripped tail. It may be possible that the signal due to the supernovae exploding in the tail is simply overcome by that of the CRe coming from the disk \citep[see][]{Ignesti_2021,Ignesti_2022d}. In order to rule out the presence of extra-planar star formation, future optical and UV studies of the stripped tails are required \citep[e.g.,][]{Poggianti2019,Giunchi_2023,Waldron_2023}. The CRe re-acceleration, instead, by extending the lifetime of the CRe beyond their supposed radiative time (Equation \ref{cool_RW}), should act on the flux density profile by extending it with an additional  component with a characteristic uniform surface brightness. These re-accelerated tails have been observed more and more frequently in radio galaxies thanks to LOFAR, and it has been explained as the consequence of the CRe being `gentle re-accelerated' by the ICM turbulence for $t>t_{\text{acc}}$, where $\tau_{\text{acc}}>100$ Myr is the re-acceleration timescale \citep[e.g.,][]{deGasperin_2017, Botteon_2021,Ignesti_2022b,Edler_2022}. Consequently, not observing low-brightness tails in these star-forming galaxies may imply that the re-acceleration is not efficient enough to compensate the energy losses, either due to synchrotron losses at 144 MHz or adiabatic expansion of the relativistic plasma (i.e., $t_{\text{rad}}<t_{\text{ad}}<t_{\text{acc}}$). These broad constraints might suggest that, for those CRe emitting at 144 MHz, $t_{\text{rad}}<100$ Myr, thus, according to Equation \ref{cool_RW}, that the magnetic field is, at least, $\sim$7 $\mu$G.
\subsubsection{Implications for the stripped ISM}
Our results can provide constraints for the velocity of the stripped material due to RPS and, thus, they can help in the study of the evolution of the stripped clouds outside of the stellar disks \citep[e.g.,][]{Sparre_2020,Tonnesen_2021,Farber_2022}. Albeit the order of magnitude is consistent with the previous numerical simulations, they predict that the cloud should decelerate, as a consequence of the mixing with the ICM, within $\sim100$ kpc from the disk \citep[e.g.,][and references therein]{Tonnesen_2021}. In this context, our model shows that the average velocity of the clouds within the first tens of kpc is relatively constant, thus the ICM mixing, and the consequent deceleration, have not yet significantly affected the clouds. Similarly, the observed spectral steepening indicates that, at least within the first tens of kpc from the stellar disk, the radio plasma can cool down undisturbed. This implies that the radiative time is shorter than the timescales of those processes that would eventually lead to the destruction of the stripped clouds, such as adiabatic expansion or mixing. Therefore, this piece of evidence tentatively constrains the order of magnitude of the stripped ISM clouds lifetime outside of the disk to be, at least, of the order of tens of Myr. 

\subsection{A potential constrain on the 3D galaxy motion}
\begin{table*}[t]
    \centering
    \begin{tabular}{cccccc}
    \toprule
    ID& $V_{\text{sky}}$ & $V_{\text{los}}$ &$V_{\text{tot}}$ &$n_e$ & $P_{\text{Ram}}$ \\
    &[km s$^{-1}$]&[km s$^{-1}$]&[km s$^{-1}$]&[$\times10^{-3}$ cm$^{-3}$]&[$\times10^{-11}$ erg cm$^{-3}$]\\
    \midrule
   1&429$\pm$83&-935&1029$\pm$34&0.07$\pm$0.01&0.14$\pm$0.01\\
3&215$\pm$32&-1355&1372$\pm$5&0.81$\pm$0.03&2.91$\pm$0.1\\
4&325$\pm$38&-64&331$\pm$37&0.99$\pm$0.04&0.21$\pm$0.05\\
6&160$\pm$32&-961&975$\pm$5&0.19$\pm$0.01&0.34$\pm$0.01\\
7&251$\pm$44&-928&962$\pm$11&0.31$\pm$0.01&0.56$\pm$0.02\\ 
    \midrule
    \end{tabular}
    \caption{From left to right: Galaxy ID (see Table \ref{sample_tab}); Best-fit $V_{\text{sky}}$; $V_{\text{los}}$ derived from galaxy redshift (see Table \ref{sample_tab}); Total velocity computed as $V_{\text{tot}}=\sqrt{V_{\text{sky}}^2+V_{\text{los}}^2}$; Electron density at the projected clustercentric distance (Figure \ref{xcop}); Ram pressure computed as $P_{\text{Ram}}=1.19\mu m_p n_e V_{\text{tot}}^2 $.}
    \label{results}
\end{table*}
In the RPS framework, the radio plasma, together with the ISM, is being displaced by the ram pressure wind. Consequently, the stripped plasma bulk velocity, at least within the first tens of kpc from the disk, should be comparable with the galaxy velocity with respect to the ICM. The observed flux density profile of a radio tail should keep track of this information, and so we speculate that the velocity estimated by our model can be used to constrain the galaxy velocity in the cluster. Specifically, the modulus of the projected velocity $V$ would represent a lower limit\footnote{Due to our assumptions on the magnetic field, see Section \ref{model}.} of the projected galaxy velocity $V_\text{sky}$ at which the galaxy is moving with respect to the ICM along the plane of the sky. The velocity component along the line of sight can be derived from the galaxy redshift $z$ by following the method described in \citet[][]{Davis_2014} to compute the peculiar velocities: 
\begin{equation}
V_{\text{los}}=c\left(\frac{z-z_{cl}}{1+z_{cl}}\right) \text{.}
\end{equation} 
 Therefore, the galaxy total 3D velocity would be $V_{\text{tot}}=\sqrt{V_\text{sky}^2+V_\text{los}^2}$. Following this approach, we estimate the galaxies' 3D velocity by adopting the velocity measured at 400 MHz as $V_{\text{sky}}$. The resulting total velocities, summarized in Table \ref{results}, span from 1 to 2$\times\sigma_{cl}$. The 3D velocities, with the only exception of galaxy $\#3$, seems to be dominated by the line-of-sight velocity. 
  
As a \textit{proof-of-concept}, in the following we explore the potentiality of using the CRe bulk velocity as indicator of the galaxy velocity along the plane of the sky.
\subsubsection{Comparing different estimators of $V_{\text{sky}}$}
\label{velocities_disc}
\begin{figure}
\includegraphics[width=\linewidth]{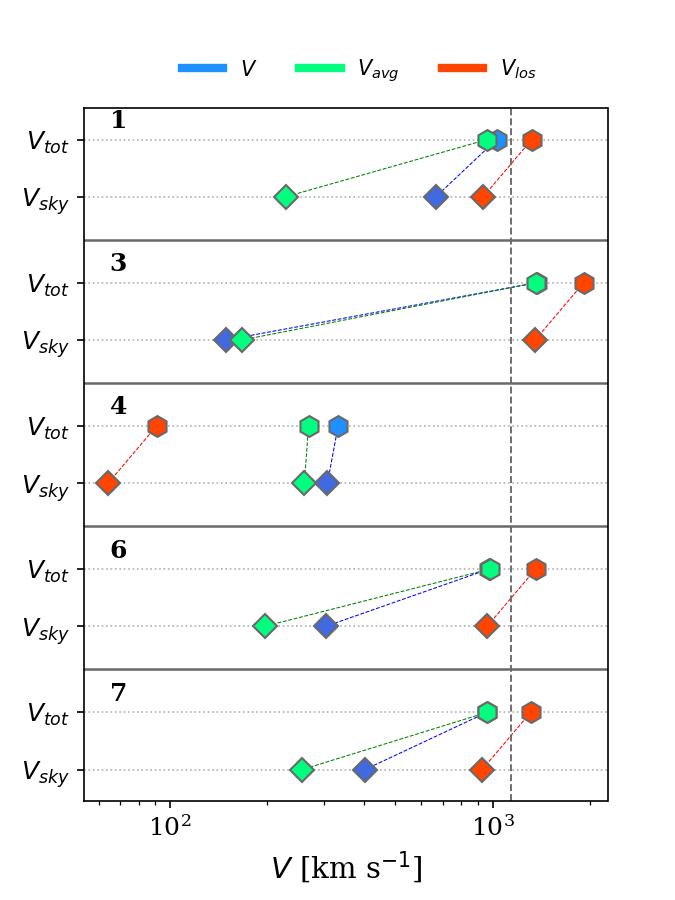}
\caption{\label{velocities_plot}
 For each galaxy: comparison between the $v_{\text{sky}}$ derived with 3 different methods (lower row, diamonds) and the corresponding total velocities (upper row, hexagons), connected by the dashed lines. The 3 methods are: the best-fit velocity $V$ derived from the 400 MHz profile (blue), $v_{\text{sky}}=v_{\text{los}}$ (red), and $v_{\text{sky}}=v_{\text{avg}}$ (green). The vertical dashed line points out the value of $\sigma_{\text{cl}}$.}
\end{figure}

First, we compare the velocity inferred from the radio tails' properties with other methods previously adopted in the literature. This comparison should be taken carefully, due to the small size of our sample, and the fact that they are all part of the same cluster. Nevertheless, this exercise can provide some insights to better evaluate the limits of each method in future studies. The two main methods adopted in previous works are:
\begin{itemize}
    \item The \textit{$45^{\circ}$ approximation}: observing RPS-induced tails, at any wavelength, is evidence that the galaxy has a significant component of its velocity directed along the plane of the sky. Thus, at the zeroth-order approximation, we can assume that, at least, $V_{\text{sky}}=V_{\text{los}}$, hence $V_{\text{tot}}=V_{\text{los}}\sqrt2$. This is equivalent to assuming that the galaxy motion is inclined of $45^{\circ}$ with respect to the line-of-sight. This method has been adopted to constrain the order of magnitude of the ram pressure given the ICM density \citep[e.g.,][]{Poggianti_2019,Campitiello_2021,Bartolini_2022};  
    \item The \textit{cooling length}: this method is based on the same physical assumption of our work, i.e. that the radio tail length depends solely on the CRe bulk velocity and the radiative time. However, in this case the hypothesis is that the CRe emit all of their energy within the observed radio tail. The average CRe velocity, $V_{\text{avg}}$, is directly computed as the ratio between the total radio tail length and the radiative time. This method has been applied when the radio data did not allow a detailed sampling of the flux density decline \citep[][]{Ignesti_2021,Muller_2021}. Its strongest caveat is that the resulting velocity mostly depends on the observed (projected) tail length, that ultimately depends on the image sensitivity. 
\end{itemize}
We compute $V_{\text{sky}}$ for each galaxy by adopting the two methods described above. For the latter one, we use the distance of the last bin of each galaxy as a measure of the radio length, and a $t_{\text{rad}}\simeq2\times10^{8}$ yr derived from Equation \ref{cool_RW} under the same assumptions of our fit ( $B=B_{\text{min}}$, $z=0.08012$, and $\nu=144$ MHz). The results are shown in Figure \ref{velocities_plot}, where, for each galaxy, we report the 3 different estimates of $V_{\text{sky}}$ (bottom row, diamonds), and the corresponding $V_{\text{tot}}$ (upper row, hexagons).

The cooling length method generally produces the lowest values of $V_{\text{sky}}$ compared with the two other methods, with the only exception being galaxy $\#3$ for which $V_{\text{avg}}\simeq V$. This result suggests that assuming the CRe have exhausted their energy within the observed length may not be correct, and not taking into account the intrinsic curvature of the radio emissivity with the time (distance) leads to underestimating $V_{\text{sky}}$ by a factor $\leq3$, due to the nonlinearity of the $t_{\text{rad}}-\nu$, and hence $t_{\text{rad}}-D$, relation. Concerning the other method, the results are  varied. For galaxy $\#4$, the $V_{\text{los}}$ is extremely low (64 km s$^{-1}$), thus suggesting that using the $45^{\circ}$ approximation may have lead to underestimate the velocity. In this case, the independent estimate provided by the radio emission permits to constrain a more realistic value of the velocity. In the other galaxies the velocity estimates are more similar. As mentioned above, this could be due to the sample bias due to the physical properties of the cluster and the galaxies within. However, albeit the results are similar, computing the velocity on the basis of the radio emission decline permits us to investigate also the geometry of the galaxy-wind interaction (see Section \ref{discussione angoli}), which is not possible otherwise. Thus this methodology would provide an advantage with respect to the $45^{\circ}$ approximation.

\subsubsection{Measuring the effective ram pressure}
\begin{figure}
    \centering
    \includegraphics[width=\linewidth]{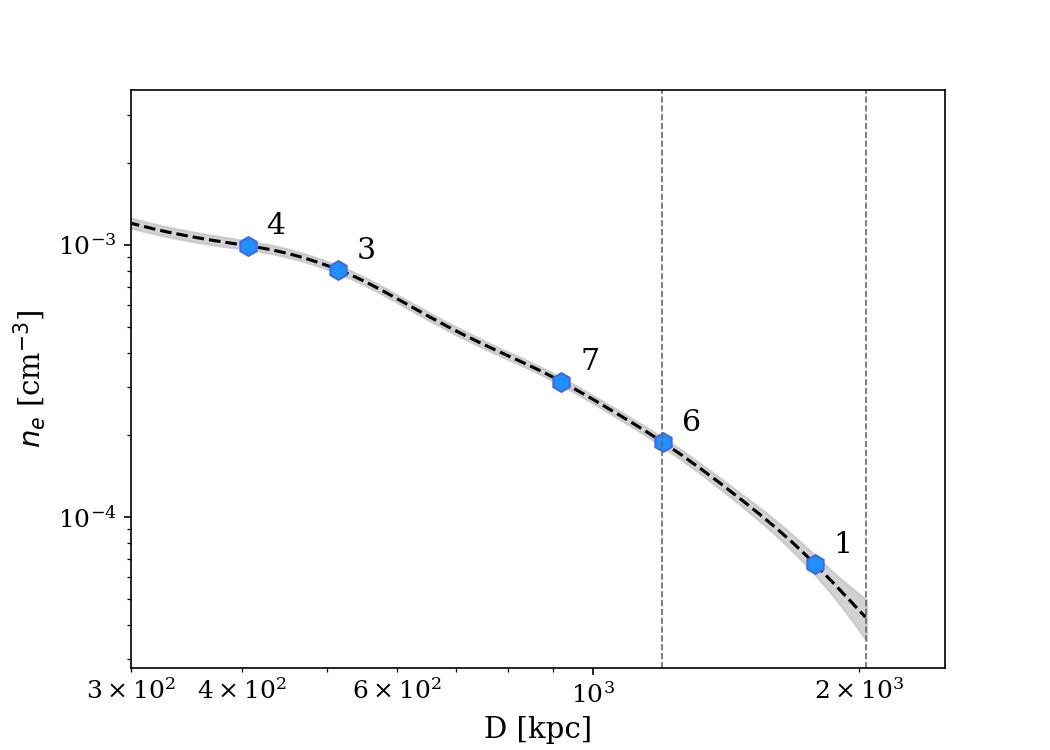}
    \caption{Electron density profile reported in X-COP. The blue points mark the galaxy positions, and the vertical dashed lines indicate, respectively, $R_{500}$ and $R_{200}$.}
    \label{xcop}
\end{figure}

Given the galaxy velocity and position in the cluster, the corresponding ICM ram pressure can be computed as  $P_{\text{Ram}}=\rho_{\text{ICM}} V_{\text{tot}}^2$, where $\rho_{\text{ICM}}=1.19\mu m_p n_e$ and $\mu$, $m_p$ and $n_e$ are, respectively, the average molecular weight, the proton mass, and the electron density \citep[e.g.,][]{Gitti_2012}. To compute the latter, we use the azimuthally-averaged electron density profile\footnote{{\tt https://dominiqueeckert.wixsite.com/xcop/a2255}} measured by the X-COP survey \citep[][]{Ghirardini_2019} to evaluate the proper $n_e$ at the projected clustercentric distance of each galaxy (Figure \ref{xcop}). The corresponding uncertainties are derived by propagating the error on the fit and the uncertainty on $n_e$. We report the results in Table \ref{results}. The caveats, limitations, and assumptions of this method are discussed in Section \ref{caveats}.

In general, the resulting values of $P_{\text{Ram}}$ lie in the $10^{-12}-10^{-11}$ erg cm$^{-3}$ range, which is in line with the previous predictions \citep[][]{Roediger2005,Bruggen_2008,Jaffe2018,Boselli_2022}. 

\subsubsection{Computing the disk-wind angle}
\label{discussione angoli}
\begin{figure}
    \centering
    \includegraphics[width=.8\linewidth]{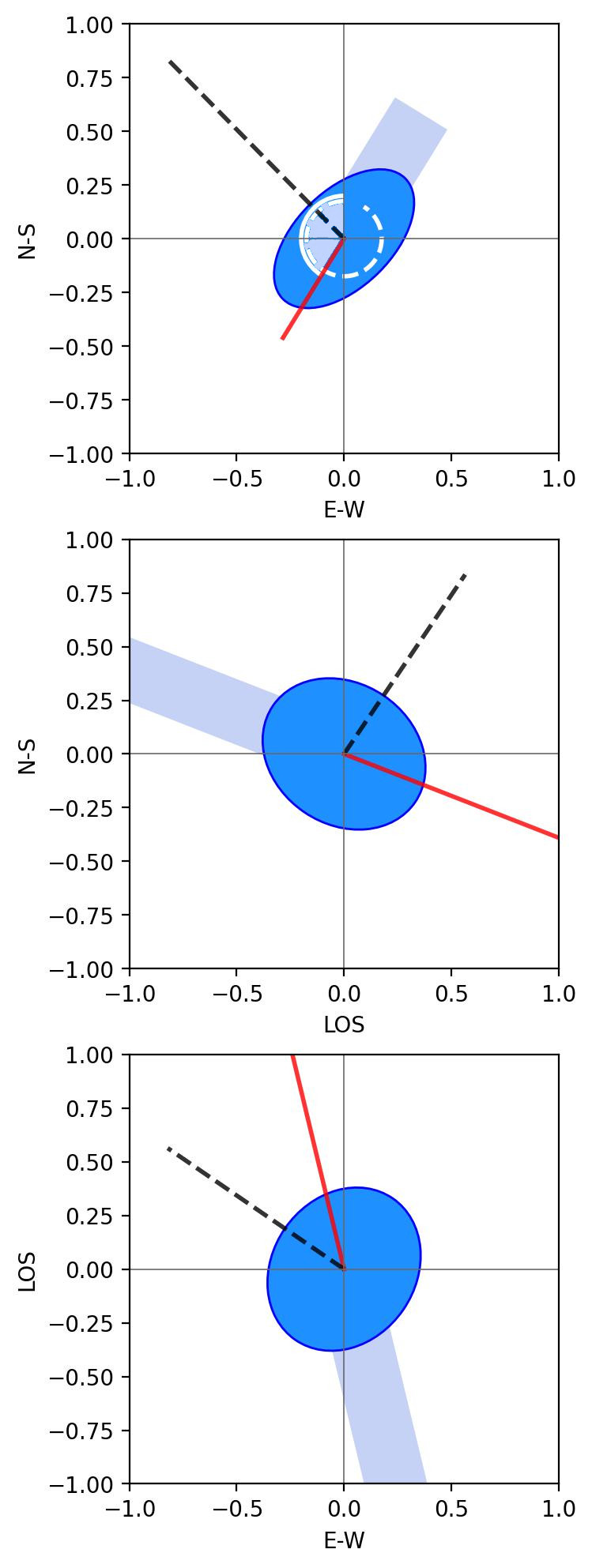}
    \caption{Example of the 3D projections in the cartesian system for galaxy $\#1$. The blue disk and segment represent the stellar disk and the observed radio tail. The red and black vectors are, respectively, the projections of $\hat{v}$ and $\hat{n}$. From top to bottom: projection along the LOS (which corresponds to the observed projection), along the E-W axis, and along the N-S axis. In the top panel we show the $\theta_V$ (white, filled) and the $\theta_{\text{tail}}$ (white, dashed) angles.} 
    \label{projection}
\end{figure}
Measuring the two components of the 3D velocity independently allows us to constrain the inclination of the galaxy disk with respect to the direction of its motion, which corresponds to the geometry of the ram pressure wind with respect to the stellar disk. Indeed several studies have shown that the evolution of ram pressure-stripped galaxies can be affected by the inclination of the ram pressure wind, which is opposite to the galaxy motion, and the stellar disk \citep[e.g.,][]{Roediger_2006, Jachym_2009, Bekki_2014,Steinhauser2016, Farber_2022, Akerman_2023}. The derivation of the disk-wind angle has been done previously for individual galaxies \citep[e.g.,][]{Vollmer_2012,Merluzzi2013}. Here we derive the disk-wind angles for the 5 galaxies, to increase the number of systems with this crucial information, and show the geometrical model that takes advantage of the results of our radio analysis. 

\begin{table}[]
    \centering
    \begin{tabular}{cccccc}
    \midrule
        ID &$\theta_{\text{tail}}$& $\phi_{\text{disk}}$&$PA$&$\Theta$&$\phi_{\text{V}}$ \\
    \midrule
1 & 328.38 & 64.3 & 134.5 & 76.85 & 35.5 \\
3 & 333.69 & 52.1 & 153.1 & 52.43 & 6.27 \\
4 & 131.22 & 41.5 & 58.6 & 39.45 & 78.2 \\
6 & 227.18 & 55.8 & 178.5 & 43.68 & 17.55 \\
7 & 351.0 & 90.0 & 154.2 & 96.6 & 23.43 \\

    \midrule
    \end{tabular}
    \caption{From left to right: galaxy ID; North-to-East angle between the galaxy center and the tail direction in the sky}; inclination and position angle as reported on HyperLeda; angle between the polar axis and the velocity versor; inclination of the velocity with respect to the line-of-sight. The values are in units of degree. We excluded galaxy $\#5$ due to the uncertainties on the fit.
    \label{angoli_disco_vento}
\end{table}
We compute the disk-wind angle, $\Theta$, starting from our estimates of $V_{\text{sky}}$ and $V_{\text{los}}$. We adopt a reference system in which the x- and y- axis are on the plane of the sky, respectively along the East-West and the North-South directions, and the z-axis coincides with the line-of-sight and points towards the observer. For each galaxy, we use the inclination of the stellar disk with respect to the \textit{line-of-sight}, $\phi_{\text{disk}}$, and its position angle, $PA$, reported in the HyperLeda\footnote{\url{http://leda.univ-lyon1.fr/}} database \citep[][]{Makarov_2014}. With them, we define the polar versor $\hat{n}$ in a cartesian coordinate system centered on the galaxy with the 3 components aligned, respectively, along the directions north-south, east-west, and the \textit{line-of-sight} (Figure \ref{projection}). Similarly, with $V_{\text{sky}}$ and $V_{\text{los}}$, where $V_{los} > 0$ indicates a galaxy moving away from the observer, and the direction of the motion, which we estimate as $\theta_V=\theta_{\text{tail}}-\pi$ where $\theta_{\text{tail}}$ is the North-to-East angle between the galaxy center and the direction of the tail in the sky, we define the velocity versor $\hat{v}$. Therefore,
\begin{equation}
\begin{aligned}
    \hat{n}=&[\cos{(PA)}\sin{(\phi_{\text{disk}})},\\
    &\sin{(PA)}\sin{(\phi_{\text{disk}})},\\
    &\cos{(\phi_{\text{disk}})}]\\
    \hat{v}=&[V_{\text{sky}}\cos{(\theta_V+\pi/2)},\\
    &V_{\text{sky}}\sin{(\theta_V+\pi/2)},\\
    &-V_{\text{los}}]/V_{\text{tot}}
    \end{aligned}
\end{equation}
Then the disk-wind angle can be computed as the angle between $\hat{v}$ and $\hat{n}$, hence:
\begin{equation}
    \Theta=\text{arccos}\left(\hat{v}\cdot\hat{n} \right)\text{ .}
\end{equation}
In this reference system, $\Theta=90$ indicates a wind impacting the galaxy edge-on, whereas $\Theta=0$ or $\Theta=180$ indicates that the galaxy is facing the wind face-on. We can measure also the inclination of the wind with respect to the \textit{line-of-sight}, $\phi_{\text{V}}$, as:
\begin{equation}
    \phi_{\text{V}}=\text{arctan}\left(\frac{V_{\text{sky}}}{V_{\text{los}}} \right)\text{ .}
\end{equation}
We report the results for each galaxy in Table \ref{angoli_disco_vento}. 

\begin{figure}
   \centering
   \includegraphics[width=.85\linewidth]{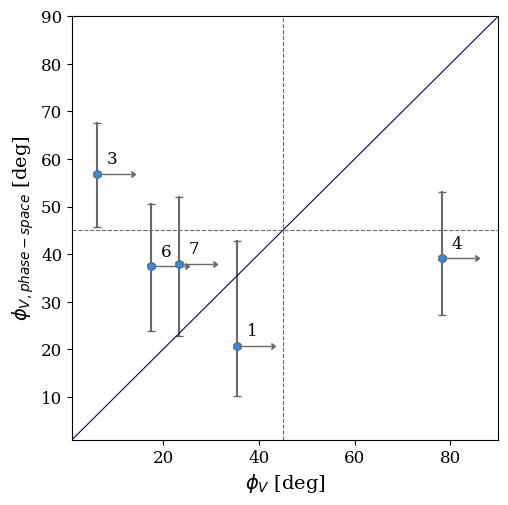}
   \caption{A comparison between the values of $\phi$ inferred in our work (x-axis) vs. the prediction based on the phase-space analysis (y-axis). The vertical errorbars indicate the first and third quartile of the $\phi$ distributions predicted by the phase-space analysis., The dashed lines separates the two regimes $\phi_{\text{V}}>45^\circ$ ($V_{\text{sky}}>V_{\text{los}}$) and $\phi_{\text{V}}<45^\circ$ ($V_{\text{sky}}<V_{\text{los}}$), and the continuous line indicates the 1:1 identity. }
   \label{rory}
\end{figure}

We compare the values of $\phi_{\text{V}}$ inferred by our analysis vs. those derived from an independent method that is the phase-space analysis presented in  \citet[][Section 3]{Bellhouse_2021}. The probable angle of a galaxy’s velocity between \textit{line-of-sight} and plane of the sky is estimated by comparing the observed $R_{CL}/R_{200}$ and $V_{\text{los}}/\sigma_{cl}$ with a phases-space diagram composed by stacking the galaxies of 42 simulated galaxy clusters observed from random angles \citep[e.g.,][for different applications of the phase-space analysis]{Smith_2022,Canducci_2022,Smith_2022b,Awad_2023}. This method provides us with a distribution of $\phi_{\text{V}}$ for each position in the phase-space diagram (i.e., for each couple of $R_{CL}/R_{200}$ and $V_{\text{los}}/\sigma_{cl}$). In Figure \ref{rory} we report the values of $\phi_{\text{V}}$ estimated by our analysis vs. the value range inferred from the phase-space analysis. We remind the reader that, due to our assumption on the magnetic field, we can derive only a lower limit for $V$, and hence $V_{\text{sky}}$. Correspondingly, the values of $\phi_{\text{V}}$ have to be considered as lower limits for the real $\phi_{\text{V}}$. For three galaxies, $\#1$, $\#7$, and $\#6$ the two predictions are in broad agreement, in the sense that they lay in the same quadrant $\phi_{\text{V}}<45^\circ$, and for galaxy $\#1$ and $\#7$ they are actually consistent. For galaxy $\#4$ the phase-space prediction is consistent within the third quartile with $\phi_{\text{V}}>45^\circ$. Therefore, for 3(+1) out of 5 galaxies the two independent analysis are in agreement. However, we note that the phase-space analysis is a statistical tool intended for large numbers of galaxies, thus this current comparison, albeit instructive, is not conclusive.
 
\subsection{Caveats}
\label{caveats}
We summarize here the caveats, limitations, and assumptions of our model.
\begin{enumerate}
\item In order to obtain a reliable fit of the flux density decline it would be best to sample the radio tails with, at least, 3 spatial bins because the velocity fit aims to constrain the curvature of the profile. This is possible only with the correct combination of sensitivity and resolution of the radio images; 
\item The best-fit $V$ depends on the shape of the underlying emissivity spectrum, which ultimately depends on the assumptions on $\delta$ and $B$. 
Assuming a steeper CRe distribution than the one we adopt ($\delta=-2.2$) will produce a steeper initial spectral index than -0.6. In principle, the initial spectral index could be measured directly from the synchrotron spectrum within the stellar disk. However, in our case this measure may not be reliable because a low-frequency spectral index in presence of high density, star forming regions, can be flattened by ionization losses, and thus it does not reflect the real CRe energy distribution \citep[e.g.][]{Basu_2015,Ignesti_2021}. Concerning the magnetic field, due to the fact that there are no methods to reliably measure its intensity in the tails, we assumed the minimal energy loss field $B_{\text{min}}$. Assuming higher values of $B$ would result also in a steeper emissivity decline for $\nu>\nu_{\text{br}}$. The magnetic field assumption defines the conversion from radiative time to projected distance. Using $B_{\text{min}}$ entails that we are working under the favorable hypothesis of maximum CRe radiative time, hence, for a given velocity, we are maximizing the distance that they can travel. Therefore, the $V$, and hence $V_{\text{sky}}$, derived under this assumption is a lower limit of the real velocity. Using different values of $B$ would entail shorter $t_{\text{rad}}$ (Equation \ref{cool_RW}), thus higher velocities are required to reproduce the observed projected distances. We quantify this behavior in Figure \ref{gb}, in which we show how the ratio of $t_{\text{rad}}$, and $V_{\text{sky}}\propto1/t_{\text{rad}}$ changes with respect to the case $B=B_{\text{min}}$ for different values of $B/B_{\text{min}}$. We observe that the ratio exceeds by a factor $2\times$ for $B\geq3.5\times B_{\text{min}}\simeq7.7$ $\mu$G; 
\begin{figure}
    \centering
    \includegraphics[width=\linewidth]{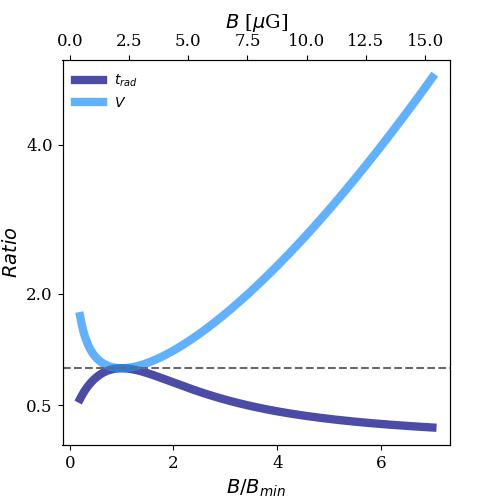}
    \caption{Expected ratio of $t_{\text{rad}}$ and $V$ with respect to the case $B=B_{\text{min}}$ for different values of $B/B_{\text{min}}$.}
    \label{gb}
\end{figure}

\item We assume that $B$ and the CRe bulk velocity $V$ are uniform along the tail. Different conditions would divert the observed flux density profiles from the prediction of pure synchrotron cooling. For instance, a magnetic field decreasing along the tail (e.g., as consequence of adiabatic expansions) could entail both a lower synchrotron emissivity but also the fact that, at a given frequency, the emission would be provided by CRe with increasing energy, and hence lower radiative times. Moreover, we assume that the CRe velocity along the tail is defined by the wind velocity, thus we neglected possible effect of the CRe stream along the magnetic field lines \citep[e.g.,][]{Armillotta_2022};
\item By fitting the emissivity spectrum directly to the observed flux density we are also inherently assuming that, in each bin, the radio-emitting plasma has the same geometrical properties, such as the volume, the curvature  along the line of sight, and the filling factor. A decreasing volume/filling factor would induce an additional decline in flux density not included by our model; 
\item The exact ram pressure, $P_{\text{Ram}}$, is derived from the azimuthally-averaged $n_e$ profile (Figure \ref{xcop}), which is computed under the assumption of spherical geometry of the ICM. To infer the appropriate $n_e$ for each galaxy we use their projected cluster-centric distance, that is a lower limit of their real distance. Moreover, the assumption of spherical geometry may not hold in a complex, merging cluster such as A2255. Therefore the values of $n_e$ reported in Table \ref{results} should be considered upper limit of the real density, and so $P_{\text{Ram}}$.  
    
\end{enumerate}
\section{Conclusions}
In this work we present a semi-empirical model to reproduce the multi-frequency radio emission of ram-pressure stripped tails, and its application. In order to test the model, we investigate the properties of the radio tails of 7 spiral galaxies in A2255. We combined LOFAR and uGMRT observations at 144 and 400 MHz to infer the radio properties within few tens of kpc from the stellar disk. We observe a monotonic decrease in flux density associated with a spectral steepening along the stripping direction. Then we modeled the observed profiles with a semi-empirical model where the radio plasma moves with a uniform velocity $V$ along the stripping direction and cools down via synchrotron radiation. The model reproduces the observed profiles for 5 out of 7 galaxies, and constrains the projected radio plasma velocity along the tail to be of the order of 100-500 km s$^{-1}$. This result confirms the qualitative scenario built up over the years in the literature, and provides the first estimate of the radio plasma bulk velocity. Moreover, observing a monotonic spectral steepening entails that, at least within the first $\sim30$ (projected) kpc from the stellar disk, the radiative time, which is of the order of $\sim100$ Myr, is shorter than the adiabatic losses timescale.   

The best-fit velocity order of magnitude supports the idea that the radio plasma clouds are transported by the ram pressure winds. Therefore, we speculate that this approach, in addition to measure the CRe bulk velocity, can constrain the galaxy velocity along the plane of the sky and provide us the first estimate of the 3D velocity of these galaxies. As a \textit{proof-of-concept}, we estimate the total velocity of these galaxies with respect to the ICM to be between 300 and 1300 km s$^{-1}$. We also infer the corresponding ram pressure exerts by the ICM to be between 0.1 and 2.9 $\times10^{-11}$ erg cm$^{-13}$, and the angle between the stellar disk and the ram pressure. These results represent the first estimates of these quantities for cluster galaxies with this method, thus they could be used to constrain future studies of these systems.

The proposed model should be now tested and refined on a larger sample of RPS galaxies, and by  using multi-frequency observations spanning a wider wavelength range. It would greatly benefit from independent estimates of the extraplanar magnetic field, potentially provided by polarimetry studies. Moreover, our results should be complemented by tailored numerical MHD simulations of RPS. On the upside, the method presented in this manuscript can expand the applications of radio observations of RPS galaxies, whose availability is destined to increase in the next years with the advance of the all-sky surveys. The combination with deep radio, X-ray and optical will permit to quantitatively characterize the RPS affecting galaxies in dense environment. 
\section*{Acknowledgments}
We thank the referee for the suggestions which improved
the quality of the manuscript. This project has received funding from the European Research Council (ERC) under the European Union's Horizon 2020 research and innovation programme (grant agreement No. 833824). AI acknowledges the INAF founding program 'Ricerca Fondamentale 2022' (PI A. Ignesti). This work is the fruit of the collaboration between GASP and the LOFAR Survey Key Project team (``MoU: Exploring the low-frequency side of jellyfish galaxies with LOFAR", PI A. Ignesti). R.J.vW  acknowledges support from the VIDI research programme with project number 639.042.729, which is financed by the Netherlands Organisation for Scientific Research (NWO). I.D.R.  acknowledges support from the ERC Starting Grant Cluster Web 804208. KR acknowledges support from Chandra grant GO0-21112X. AI thanks the music of Vulfpeck for providing inspiration during the preparation of the draft.

LOFAR \citep[][]{vanHaarlem_2013} is the Low Frequency Array designed and constructed by ASTRON. It has observing, data processing, and data storage facilities in several countries, which are owned by various parties (each with their own funding sources), and that are collectively operated by the ILT foundation under a joint cientific policy. The ILT resources
have benefited from the following recent major funding sources: CNRS-INSU, Observatoire de Paris and Université d'Orléans, France; BMBF, MIWF-NRW, MPG, Germany; Science Foundation Ireland (SFI), Department of Business, Enterprise and Innovation (DBEI), Ireland; NWO, The Netherlands; The Science and Technology Facilities Council, UK; Ministry of
Science and Higher Education, Poland; The Istituto Nazionale di Astrofisica (INAF), Italy. This research made use of the Dutch national e-infrastructure with support of the SURF Cooperative (e-infra 180169) and the LOFAR e-infra group. The Jülich LOFAR Long Term Archive and the German LOFAR network are both coordinated and operated by the Jülich Supercomputing Centre (JSC), and computing resources on the supercomputer JUWELS at JSC were provided by the Gauss Centre for Supercomputing e.V. (grant CHTB00) through the John von Neumann Institute for Computing (NIC). This research made use of the University of Hertfordshire high-performance computing facility and the LOFAR-UK computing facility located at the University of Hertfordshire and supported
by STFC [ST/P000096/1], and of the Italian LOFAR IT computing infrastructure supported and operated by INAF, and by the Physics Department of Turin university (under an agreement with Consorzio Interuniversitario per la Fisica Spaziale) at the C3S  Supercomputing Centre, Italy. We thank the staff of the GMRT that made these bservations possible. GMRT is run by the National Centre for Radio Astrophysics of the Tata Institute of Fundamental Research. This research made use of Astropy, a community-developed core Python package for Astronomy \citep[][]{astropy_2013, astropy_2018}, and APLpy, an open-source plotting package for Python \citep[][]{Robitaille_2012}. This research has made use of the SIMBAD database,
operated at CDS, Strasbourg, France \citep[][]{Wenger_2000}.

\bibliographystyle{aa.bst}
\bibliography{bibi.bib}
\end{document}